\documentclass[%
 twocolumn,
showpacs,preprintnumbers,
nofootinbib,
 amsmath,amssymb,
 aps,
 showkeys,
 showpacs
]{revtex4-1}
\usepackage[utf8]{inputenc}

\usepackage{graphics}
\usepackage{bbold}
\usepackage{color}

\usepackage{cancel}
\usepackage[normalem]{ulem}

\usepackage{array}
\usepackage{booktabs}
\usepackage{amsmath}
\usepackage{graphicx}
\usepackage{subfigure}

\usepackage{amsmath}
\usepackage{slashed}

\usepackage{multirow}

\usepackage{float}

\usepackage[T1]{fontenc}

\usepackage{lineno,hyperref}

\def\be		{\begin{eqnarray}}
\def\en		{\end{eqnarray}}
\def\nen	{\nonumber\end{eqnarray}}
\def\no		{\nonumber}

\def\lt		{\left(}
\def\rt		{\right)}
\def\lq		{\left[}
\def\rq		{\right]}

\def\ds		{\ensuremath{\displaystyle}}
\def\hh		{\hspace{5 mm}}

\def\ug		{\ensuremath{&=&}}

\def\LL		{\ensuremath{\Lambda\bar{\Lambda}}}

\def\L		{\ensuremath{{\Lambda}}}

\def\ee		{\ensuremath{e^+e^-}}
\def\BB		{\ensuremath{B\bar{B}}}

\def\e		{\ensuremath{\epsilon}}
\def\N 		{\ensuremath{\mathbb{N}}}

\def\GE	    {\ensuremath{G_E^\Lambda}}
\def\GM	    {\ensuremath{G_M^\Lambda}}

\def\R	    {\ensuremath{\frac{G_E^\Lambda}{G_M^\Lambda}}}

\def\qth     {\ensuremath{q^2_{\rm th}}}
\def\qasy     {\ensuremath{q^2_{\rm asy}}}
\def\qphy     {\ensuremath{q^2_{\rm phy}}}

\def\Rasy     {\ensuremath{R_{\rm asy}}}

\def\im     {\ensuremath{{\rm Im}}}
\def\re     {\ensuremath{{\rm Re}}}

\def\R {\ensuremath{\mathbb{R}}}
\def\C {\ensuremath{\mathbb{C}}}
\def\Z {\ensuremath{\mathbb{Z}}}
\def\N {\ensuremath{\mathbb{N}}}

\definecolor{amber}{rgb}{1.0, 0.55, 0.0}
\definecolor{redmod}{rgb}{0.89, 0.0, 0.13}
\definecolor{blumod}{rgb}{0.01, 0.28, 1.0}

\def\nth 	{\ensuremath{N_{\rm th}}}
\def\nasy 	{\ensuremath{N_{\rm asy}}}
\def\bmi	{\ensuremath{\begin{minipage}}}
\def\emi	{\ensuremath{\end{minipage}}}

\def\rael	{\ensuremath{\langle r_E^\Lambda\rangle}}
\def\rae	{\ensuremath{\langle r_E\rangle}}

\def\GEq 	{\ensuremath{G_E^\Lambda}(q^2)}

\def\GMq 	{\ensuremath{G_M^\Lambda}(q^2)}

\definecolor{verde1}{rgb}{0.0, 0.96, 0.0}   	
\definecolor{verde2}{rgb}{0.0, 0.84, 0.0}	
\definecolor{verde3}{rgb}{0.0, 0.495, 0.0}	
\definecolor{verde4}{rgb}{0.0, 0.993, 0.0}	
\definecolor{verde5}{rgb}{0.0, 0.732, 0.0}	
\definecolor{verde6}{rgb}{0.0, 0.984, 0.0}	

\begin{document}

\title{%
The first exploration of the physical Riemann surfaces of the ratio $G_E^\Lambda/G_M^\Lambda$%
}
%
 \author{Alessio Mangoni}
\affiliation{%
INFN Sezione di Perugia, I-06100, Perugia, Italy}%
\author{Simone Pacetti}
\affiliation{%
INFN Sezione di Perugia and Università di Perugia, I-06100, Perugia, Italy}%
\author{Egle Tomasi-Gustafsson}
\affiliation{%
DPhN, IRFU, CEA, Université Paris-Saclay, 91191 Gif-sur-Yvette Cedex, France}%
\begin{abstract}
Recently, the BESIII experiment renewed the interest on baryon form factors by measuring the modulus and phase of the ratio $\GE/\GM$ between the electric and the magnetic $\Lambda$ form factors with unprecedented accuracy. The BESIII measurement together with older, less precise data, can be analyzed by means of a dispersive procedure based on analyticity and a set of first-principle constraints. Such a dispersive procedure shows the unique ability to determine for the first time the complex structure of the ratio knowing its modulus and phase measured by the BESIII collaboration at only one energy point. Different classes of solutions are obtained, and in all cases, the time-like and space-like behaviors show interesting properties: space-like zeros or unexpected large determinations for the phase. More data at different energies would be crucial to enhance the predictive power of the dispersive procedure and to unravel further remarkable features of the $\L$ baryon.
\end{abstract}

\maketitle

\section{Introduction}
\label{sec:intro}
The structure of baryons and hence, that of the $\Lambda$, can be studied through fundamental quantities called form factors (FFs) (see Ref.~\cite{Pacetti:2014jai} for an exhaustive review). They are Lorentz scalar functions depending on the four-momentum squared related to the baryon four-current and contain pivotal information on their dynamics, as well as on some static properties.
\\
In particular the $\LL\gamma$ vertex is described by two independent FFs, usually represented by the so-called Sachs electric and magnetic FFs $G_E$ and $G_M$~\cite{sachs}. In the Breit frame, which is the reference system where there is no energy exchange, i.e., where the four-momentum of the virtual photon coincides with its three-momentum, the Sachs FFs represent the Fourier transforms of the electric charge and magnetic momentum spatial distributions of the $\Lambda$ baryon.
\\
The study of their ratio represents a crucial point, since the relative phase between the electric and magnetic FFs of baryons, i.e., ${\rm arg}\lt\GE/\GM\rt$, is directly related to their polarization.
\\
The main goals of the present study are the definition and the exploitation of a dispersive approach, based on first principles and theoretical constraints, to analyze the available data on the ratio $\GE/\GM$. A unique and novel outcome of such analysis is the complete disclosure of the complex structure of the FF ratio. In fact, besides the modulus and the phase, the procedure is able to establish the specific determination of the phase itself, i.e., the Riemann surface to which the physical values of the ratio do belong.
\section{The special case of $\Lambda$ baryons}
\label{sec:lambda}
\begin{figure}[H]
		\includegraphics[width=\columnwidth]{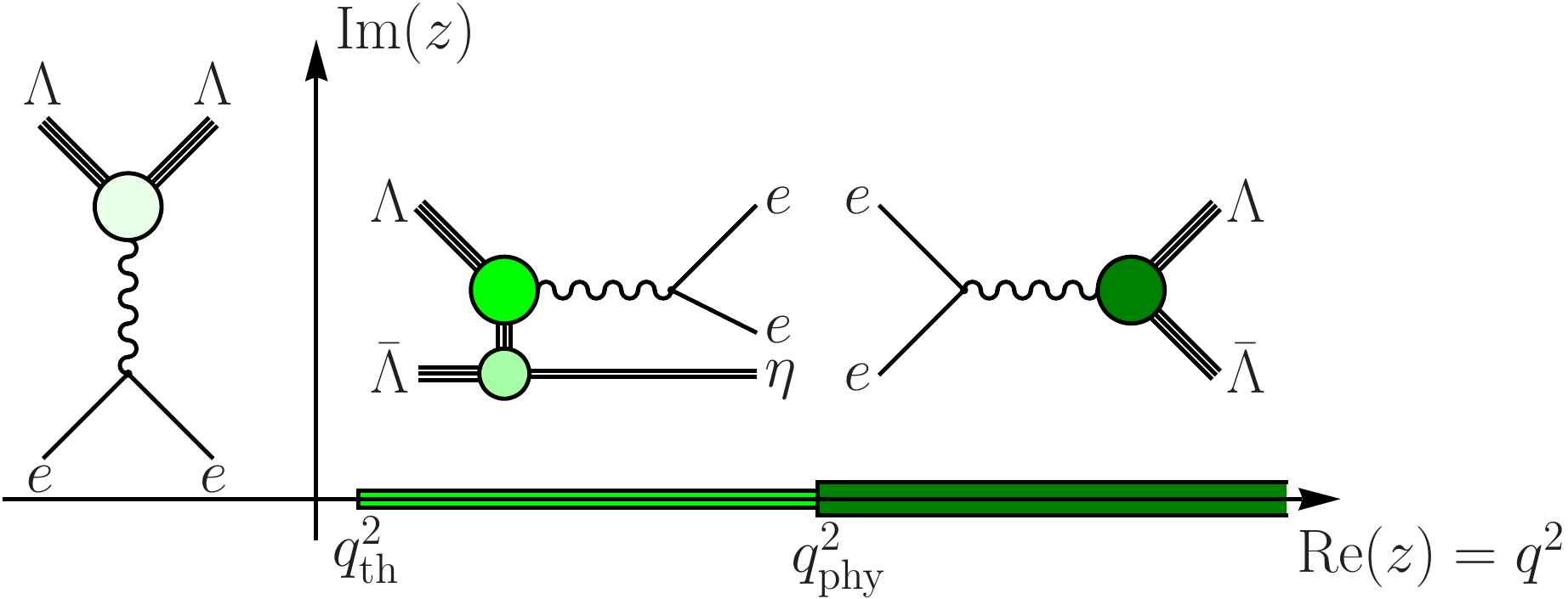}
		\caption{\label{fig:complex-plane}Schematic representation of $z=q^2$-complex plane, with the Feynman diagrams, at the lowest order, of the reactions that allow to extract FF values in each kinematic region.}
\end{figure}

The \L\ FFs, like all baryon FFs, are Lorentz scalar functions of the four-momentum transfer squared  $q^2$, analytic in the $q^2$-complex plane with a branch cut along the positive real axis, from the theoretical threshold, $q^2=\qth$ up to infinity. The theoretical threshold corresponds to the mass squared of the lightest hadronic channel that can couple with the virtual photon and with the \LL\ state. In the case of the nucleons, such a hadronic state is that of the two charged pions $\pi^+\pi^-$ and hence the theoretical threshold is $(2M_\pi)^2$, being $M_\pi$ the charged pion mass. 
\\
However, since the \L\ baryon and hence also the \LL\ system have isospin zero, the theoretical threshold of the \L\ FFs is $\qth=(2M_\pi+M_{\pi^0})^2$. Indeed the lightest isoscalar, i.e., having isospin zero, hadronic state is that of three pions: $\pi^+\pi^-\pi^0$.
\\
The only portion of the $q^2$-complex plane that is experimentally accessible is the real axis, indeed, only real values of $q^2$ can be tested in laboratories. 
\\
Due to Hermitian properties of the electromagnetic current operator of the baryon vertex $\LL\gamma$, the FFs that parametrize such a current operator are real at real values of $q^2$ belonging to the analyticity domain, namely $q^2\le\qth$, and in general verify the Schwarz reflection principle: $G(q^{2*})=G^*(q^2)$, with $G=\GE,\GM$. It follows that, on the edges of the branch cut $(\qth,\infty)$, they are complex, having a non-vanishing imaginary part. In fact, at $z=q^2+i\e$, with $q^2>\qth$ and $\e\to 0^+$, the imaginary part is
\be
\im[G(z)]
\ug \frac{G(z)-G^*(z)}{2i}
= \frac{G(q^2+i\e)-G^*(q^2+i\e)}{2i}
\no\\
\ug  \frac{G(q^2+i\e)-G(q^2-i\e)}{2i}\not=0\,,
\nen 
where the inequality follows from the discontinuity of the FFs across the branch cut. 
\\
In summary, FFs are real for real values of $q^2$ below the theoretical threshold, i.e., at $q^2\le \qth$, and hence in the whole space-like region, $q^2\le 0$, together with the time-like interval $(0,\qth)$; while they are complex in the remaining part of the time-like region, namely for $q^2\ge \qth$.
\\
In principle, even in an ideal laboratory without any experimental limits, FFs would not be completely accessible in the entire kinematic region, that is $q^2\in(-\infty,\infty)$.
\\
Form Factor values can be extracted from specific observables in each kinematic region.
\begin{itemize}
\item In the space-like region, $q^2\le 0$, \GE\ and \GM\ are real and their values can be extracted by the differential cross section of the scattering reaction $e^-\Lambda\to e^-\Lambda$. The Feynman diagram of the reaction is shown on the left side of Fig.~\ref{fig:complex-plane}, laying in the corresponding kinematical region.
\item In the time-like range $q^2>\qphy$, being $\qphy=(2M_\Lambda)^2$ the physical or production threshold, where the FFs have complex values, their moduli can be extracted from the differential cross section of the annihilation processes $\ee\leftrightarrow\LL$. Also in this case, the Feynman diagram of the reaction positioned in the corresponding kinematical region is shown on the right side of Fig.~\ref{fig:complex-plane}. Moreover, the relative phase $\arg(\GE/\GM)$ can be extracted from the measurement of the polarization of one of the outgoing baryons, along the direction orthogonal to the scattering plane.
\item In the so-called \emph{unphysical region}, i.e., between the theoretical and the physical threshold, at $\qth<q^2\le \qphy$, moduli of the FFs, in principle, can be extracted from the differential cross section of the reaction $\LL\to\eta\ee$, where the $\eta$ meson is emitted by one of the initial baryons. The corresponding Feynman diagram is shown in the middle of Fig.~\ref{fig:complex-plane}. The possibility of using this process to experimentally access the unphysical region has been deeply investigated in the case of proton FFs, for the corresponding reaction $p\bar p\to\ee\pi^0$~\cite{Re65,egle-maas}.
\end{itemize}
The present unavailability of stable target and, a fortiori, of stable beams of \L\ baryons, prevents the possibility of making scattering and \LL\ annihilation experiments. It follows that, \L\ FFs can be measured neither in the space-like, nor in the unphysical region, as it can be done in the case of stable baryons, namely for proton and neutron.
\\
 On the other hand, the weak decay $\L\to p\pi^-$, commonly exploited to identify the \L\ baryon itself thanks to the charged hadrons in the final state, has the crucial advantage to be \emph{self-analyzing}, meaning that the \L\ polarization can be extracted from the angular distribution of the final proton. In the light of this, it is possible to access the relative phase $\arg(\GE/\GM)$ without any need of direct polarization measurements that would require inserting polarimeters in the main detector, by worsening consequently its general tracking performances.
\section{Polarization observables}
\label{sec:polarization}
It is well-known~\cite{rekalo} that, even though the initial leptons are unpolarized, the baryon $B$ and the anti-baryon $\bar B$ produced by the annihilation reaction $\ee\to\BB$ are polarized along the direction orthogonal to the scattering plane, namely the plane containing the three-momenta of the initial and final particles in the \ee-center of mass frame. This \emph{spontaneous} polarization is due to the complex nature of the FFs and hence it can be used to \emph{measure} their complexity. In particular, the $y$-component of the polarization vector of the out-going baryon $B$, defining the $xz$-plane as the scattering plane, has the expression~\cite{rekalo}
\be
\mathcal{P}_y=
-\frac{\sqrt{\frac{q^2}{4M_B^2}}\frac{|\GE|}{|\GM|}  \sin(2\theta) \sin\lt\arg\lt\frac{\GE}{\GM}\rt\rt}
{\frac{q^2}{4M_B^2} \lt 1+\cos^2(\theta)\rt+
 \frac{|\GE|^2}{|\GM|^2}\sin^2(\theta)}\,,
 \label{eq:Py}
\en
where $q$ is the transfer four-momentum, $M_B$ is the baryon mass and $\theta$ is the scattering angle in the \ee\ center of mass frame. It is evident that the $y$ polarization vanishes in case of relative reality of FFs, i.e., if the relative phase $\arg(\GE/\GM)$ is an integer multiple of $\pi$ radians.
\\
On the other hand, the polarization is maximum at
\be
\cos(\theta)=\pm \sqrt{\frac{ {q^2}/{(4M_B^2)} +
  {|\GE|^2}/{|\GM|^2}}{ 3 q^2/(4M_B^2)+
 {|\GE|^2}/{|\GM|^2}}}\,,
\nen
close to the production threshold, i.e., at $q^2\simeq (2M_B)^2$, and assuming $|\GE|\simeq |\GM|$, gives $\cos(\theta)\simeq \pm 1/\sqrt{2}$, and hence: $\theta\simeq \pi/4+k\pi,3\pi/4+ k\pi$, with $k\in\Z$. 
\\
In the light of this remark, the value of the sinus of the relative phase $\arg(\GE/\GM)$ at a given $q^2$ can be extracted from the experimental data on the polarization component $\mathcal{P}_y$, measured at that $q^2$ and at any $\theta$. Indeed, the relative phase, like the FFs themselves, depends solely on the four-momentum transferred squared. 
\\
Finally, it is interesting to notice that, the knowledge of the sinus does not allow to have any clue on the determination of the relative phase whatever information it does contain.
\section{Data}
\label{sec:data}
\begin{figure}
		\includegraphics[width=.33\columnwidth]{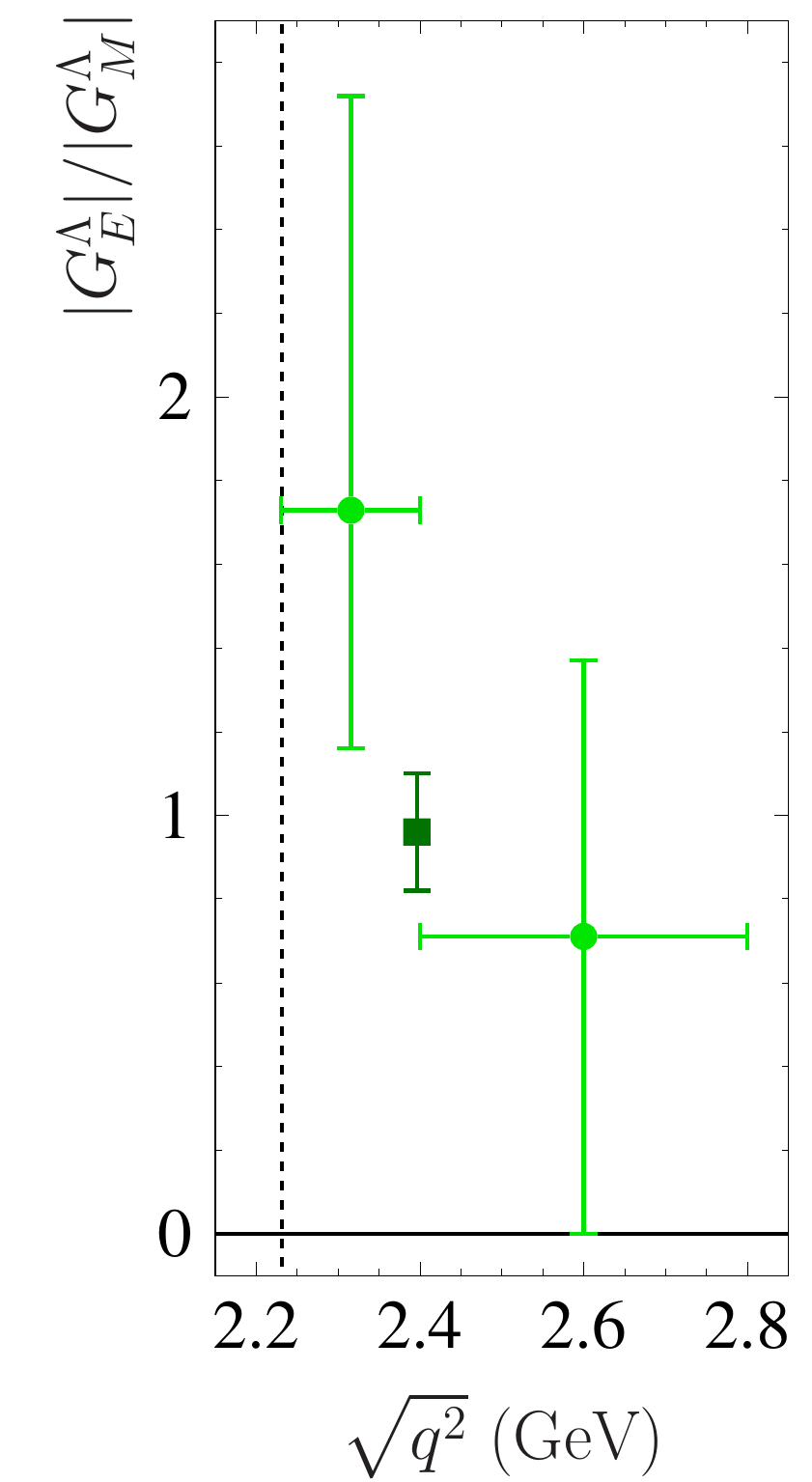}\includegraphics[width=.33\columnwidth]{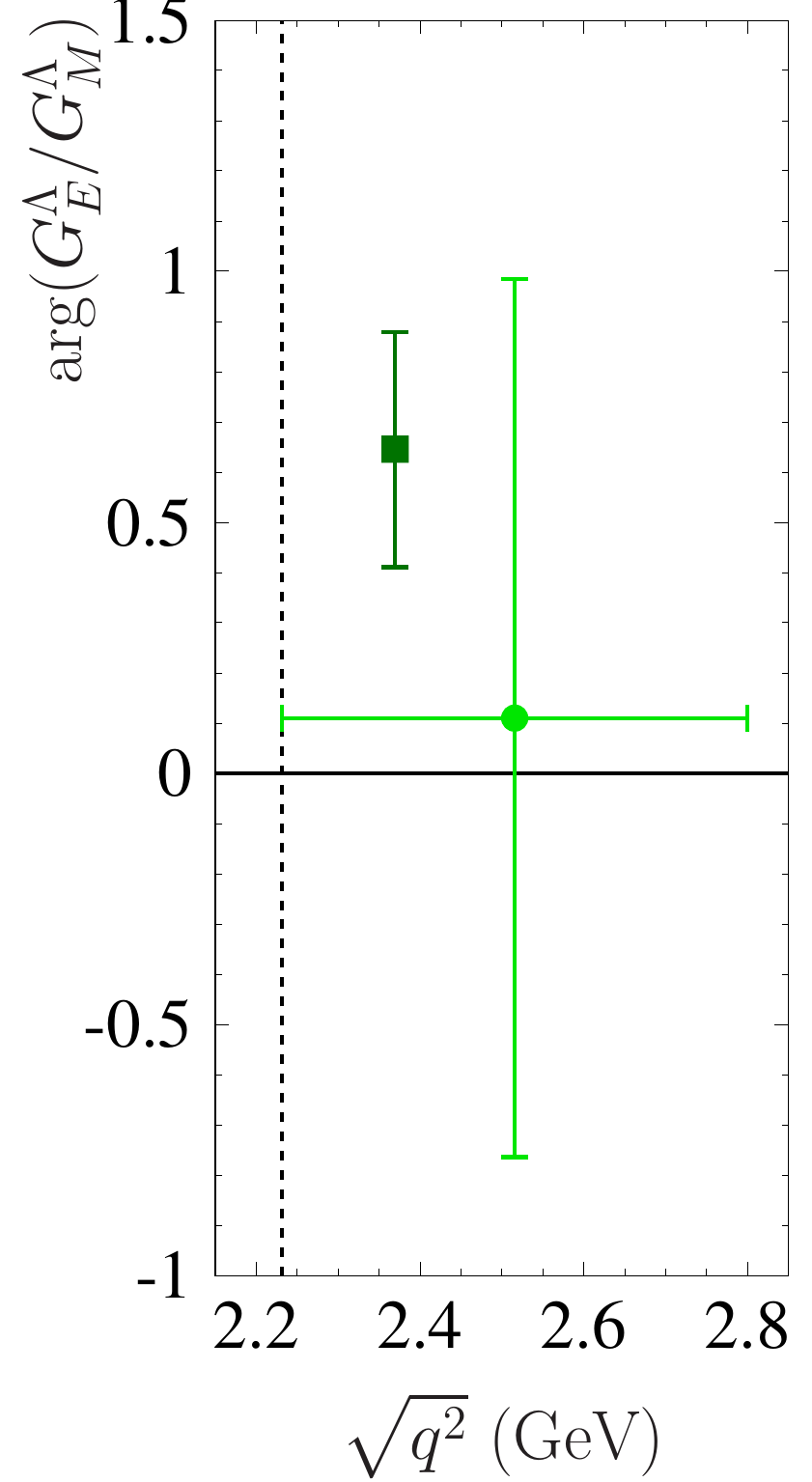}\includegraphics[width=.33\columnwidth]{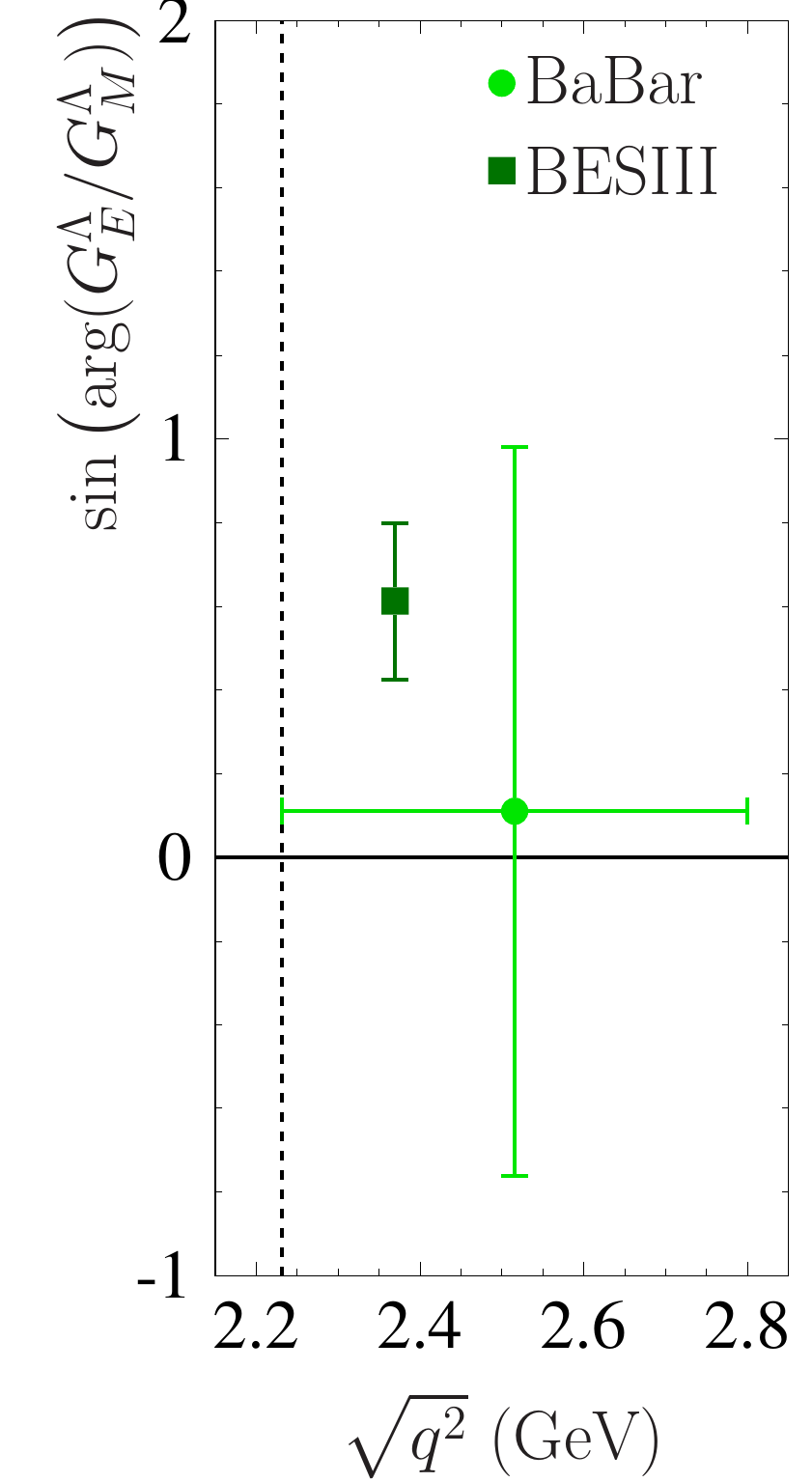}
		\caption{\label{fig:data-LL} Data on the modulus, left panel, the phase, central panel, and the sinus of the phase, right panel, of the ratio $\GE/\GM$ measured by BaBar~\cite{Aubert:2007uf}, light-green solid circles, and BESIII~\cite{Ablikim:2019vaj}, dark-green solid square, experiments.}
\end{figure}

At present, only two sets of data, both of them on the modulus and the phase of the ratio $\GE/\GM$, are available. They have been obtained by the BaBar experiment~\cite{Aubert:2007uf} in 2006 and, more recently, in 2019, by the  BESIII experiment~\cite{Ablikim:2019vaj}. The BaBar data set consists in two data points on the modulus and one point, obtained by collecting the whole statistics, on the phase of the ratio $\GE/\GM$. The low statistics of the BaBar measurement is a consequence of the initial-state-radiation technique, that has to be used at fixed-energy machines to reproduce the same observables available at a typical \ee\ machine with energy scan.
\\
The data from BESIII experiment consist in the modulus and phase of the ratio $\GE/\GM$ at a unique energy point. Nevertheless, the achieved level of accuracy is quite high because it was a direct measurement, obtained by collecting events $\ee\to\LL$ at the desired energy.
\\
All BaBar and BESIII data on the ratio $\GE/\GM$ are shown in Fig.~\ref{fig:data-LL}, in particular, from left to right, the modulus, the phase and the sinus of the phase, which actually represents the genuine experimental observable extracted from the $y$-component of the \L\ polarization vector, see Eq.~\eqref{eq:Py}.
\\
 As already highlighted, this implies that the determination of the phase%
%
%
\footnote{At a given value $x=\sin(\alpha)\in[-1,1]$ of the sinus do correspond infinite angles, namely $\alpha_k=\arcsin(x)+2k\pi$ and $\alpha'_k=-\arcsin(x)+(2k+1)\pi$, for all $k\in\Z$, with  $\arcsin(x)\in[-\pi/2,\pi/2)$.} 
%
%
is not experimentally accessible through the polarization measurement, only.
\section{On the meaning of the phase determination}
\label{sec:determiantion}
\begin{figure}[H]
\begin{center}
		\includegraphics[width=.8\columnwidth]{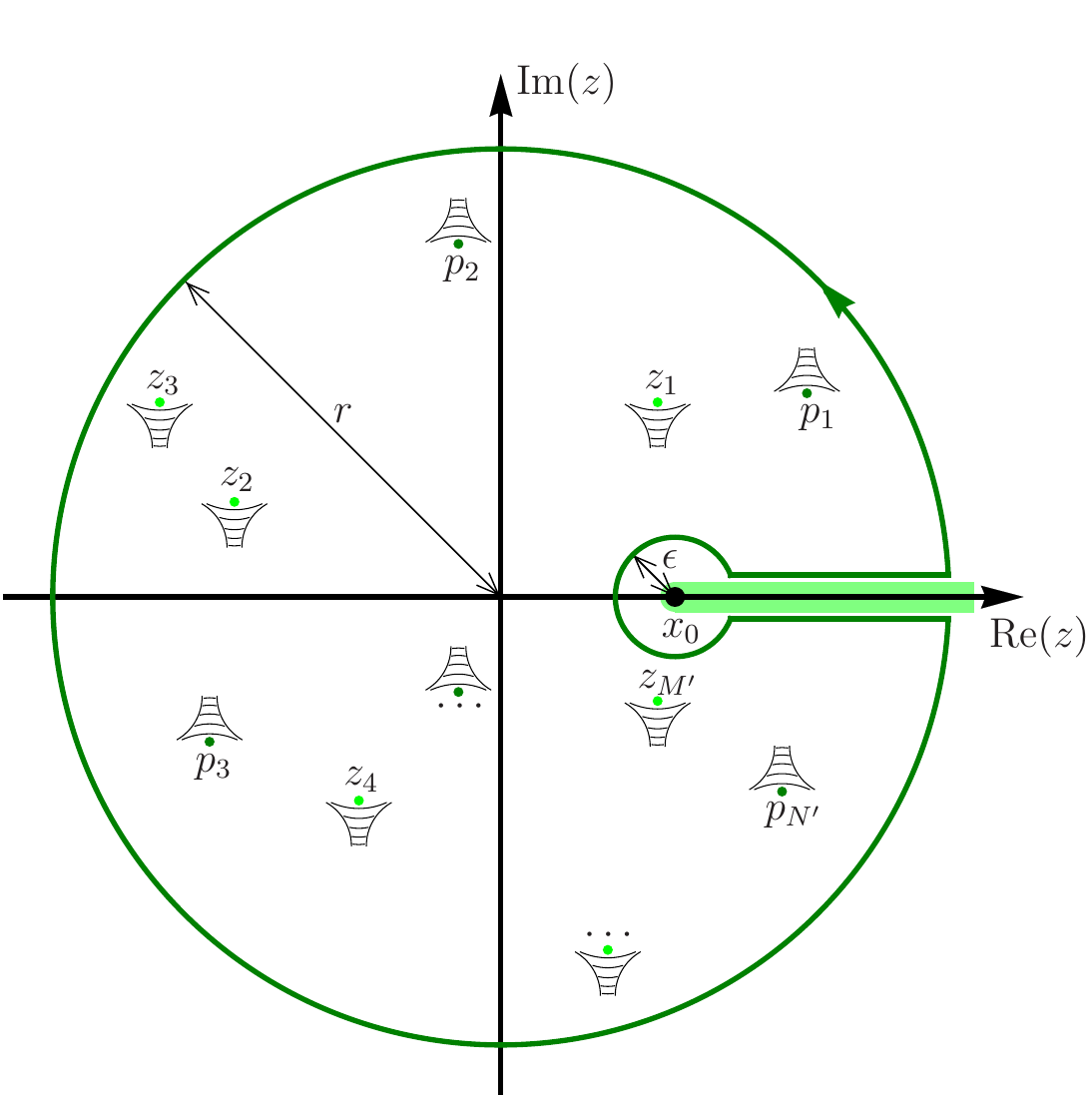}\caption{\label{fig:cauchy} The dark green curve represents the $\Gamma_{r,\e}$ integration path, while the dark and light green solid circles indicate the poles and zeroes surrounded by $\Gamma_{r,\e}$. The real and positive branch cut $(x_0,\infty)$ is highlighted by a light-green band.}
\end{center}
\end{figure}

In order to understand the meaning and hence the information that is embodied in the phase of an analytic multivalued function, and in particular in its determination, we study in some detail the following example. Consider an analytic multivalued function $R(z)$, defined in the domain $D=\{z:z\not\in (x_0,\infty)\cup \{p_j\}_{j=1}^{N'}\}$, having the real branch cut $(x_0,\infty)\subset \R$ and the set of zeros $\{z_k\}_{k=1}^{M'}\subset \C$, where $\{m_k\}_{k=1}^{M'}\subset\N$ is the set of the corresponding orders, while $\{p_j\}_{j=1}^{N'}\subset \C$ and $\{n_j\}_{j=1}^{N'}\subset\N$ are the sets of the poles and of the corresponding orders. Figure~\ref{fig:cauchy} shows a pictorial view of the domain $D$, the branch cut is highlighted in light green, while zeros and poles are indicated by dips and spikes, respectively.
\noindent
The integral of the logarithmic derivative of the function $R(z)/(2i\pi)$ over the closed path $\Gamma_{r,\e}$, shown as a dark green curve in Fig.~\ref{fig:cauchy}, which does not intersect neither poles, nor zeros, having 
\be
r&>& \max_{\ds\mathop{^{j=1,2,\ldots,N'}}_{k=1,2,\ldots,M'}}\{|p_j|,|z_k|\}\,,\no
\\
\e &<& \min_{\ds\mathop{^{j=1,2,\ldots,N'}}_{k=1,2,\ldots,M'}}\{|x_0-p_j|,|x_0-z_k|\}\,,
\nen
where $r$ and $\e$ are the radii of the circles centered in the origin and in $x_0$, gives the difference between the total numbers of zeros $M$ and poles $N$ of the function $R(z)$, lying inside the contour $\Gamma_{r,\e}$. Zeros and poles are counted as many times as their multiplicities, i.e.,
\be
\sum_{k=1}^{M'}m_k=M\,,\hh \sum_{j=1}^{N'}n_j=N\,.
\nen
It follows that
\be
\frac{1}{2i\pi}\oint_{\Gamma_{r,\e}} \frac{d\ln\lq R(z)\rq}{dz}dz=
 M-N\,,
 \label{eq:index}
\en
this is also called the \emph{argument principle} or \emph{Cauchy's argument principle}~\cite{Ahlfors}.
\\
Assuming that $z=\infty$ is an accumulation point neither of zeros nor of poles of the function $R(z)$, the result of Eq.~\eqref{eq:index} does hold also under the limits $\e\to 0^+$ and $r~\to~\infty$, so that
\be
\lim_{r\to\infty}\lim_{\e\to 0^+}\frac{1}{2i\pi}\oint_{\Gamma_{r,\e}} \frac{d\ln\lq R(z)\rq}{dz}dz=
 M-N\,.
\label{eq:limit-index}
\en
Moreover, if the function $R(z)$ has the asymptotic behavior suitable to make infinitesimal the contributions of the two arcs 
\be
\gamma_\e\ug \{z:|z-x_0|=\e e^{i\phi},\,\phi\in(\eta,2\pi-\eta),\,\eta\to 0^+\}\,,
\no\\
\gamma_r\ug\{z:|z|=r e^{i\phi},\,\phi\in(\eta,2\pi-\eta),\,\eta\to 0^+\}\,,
\nen
as $\e\to 0^+$ and $r\to\infty$, respectively, i.e.,
\be
\lim_{\e\to 0^+}\frac{1}{2i\pi}\oint_{\gamma_{\e}} \frac{d\ln\lq R(z)\rq}{dz}dz\ug 0\,,\no\\
\lim_{r\to\infty}\frac{1}{2i\pi}\oint_{\gamma_{r}} \frac{d\ln\lq R(z)\rq}{dz}dz\ug 0\,,
\nen
then the limit of the integral on the left-hand-side of  Eq.~\eqref{eq:limit-index} becomes
\be
&&\frac{1}{2i\pi}\lt \int_{x_0+i\e}^{\infty+i\e} \frac{d\ln\lq R(z)\rq}{dz}dz\right.
\no\\
&&\left.
\hspace{15mm}-\int_{x_0-i\e}^{\infty-i\e} \frac{d\ln\lq R(z)\rq}{dz}dz\rt=M-N\,.
\label{eq:limit-index2}
\en
Taking advantage from the Schwarz reflection principle~\cite{Ahlfors} $R(z)=R^*(z^*)$ and using the polar form $R(z)~=~|R(z)|e^{i\,\arg(R(z))}$, the second integral can be written in Cartesian coordinates, i.e., as the combination of its real and imaginary part and it reads
\be
\int_{x_0- i\e}^{\infty- i\e} \frac{d\ln\lq R(z)\rq}{dz}dz
\ug \int_{x_0}^{\infty} \frac{d\ln\lq R(x- i\e)\rq}{dx}dx
\no\\
\ug\int_{x_0}^{\infty} \frac{d\ln\lq R^*(x+ i\e)\rq}{dx}dx\no\\
\ug\int_{x_0}^{\infty} \frac{d\ln|R(x+ i\e)|}{dx}dx
\no\\
&&- i\int_{x_0}^{\infty} \frac{d\arg (R(x+ i\e))}{dx}dx\,.
\nen
Finally, by using this expression in Eq.~\eqref{eq:limit-index2},	 it reduces to
\be
\frac{1}{\pi}
\int_{x_0}^{\infty} \frac{d\arg (R(x+ i\e))}{dx}dx
=M-N
\,,
\nen
giving, after the integration of the left hand side,
\be
\arg(R(\infty))-\arg(R(x_0))=\pi(M-N)\,.
\label{eq:levinson}
\en
It represents the version of Levinson's theorem~\cite{levinson} which will be extensively used in the following.\\
Notice that, in the perspective of using the result of Eq.~\eqref{eq:levinson} for the ratio of FFs, which is a function of $q^2$, the infinitesimal imaginary part \e\ appearing in the argument of the function $R(z)$ has been omitted, because, as usually done, the time-like values of FFs at $q^2>\qth$ are defined as those on the upper edge of the branch cut, i.e., for $q^2>\qth$, $G_{E,M}^\Lambda(q^2)\equiv G_{E,M}^\Lambda(q^2+i\,\e)$.
\section{Dispersion relations}
\label{sec:DR}
The most powerful analytic-continuation technique, especially suitable for FFs, is represented by the so-called dispersion relations (DR's)~\cite{Walker}. The feature of being analytically continued in an extended domain containing the original one represents a crucial property of analytic functions, which is grounded on the principle of their uniqueness~\cite{Ahlfors}.
\\
There are different forms of DR's. In particular, by considering an analytic and multivalued function $f(z)$, defined in the domain $C=\{z:z\not\in (x_0,\infty)\}$, i.e., in the whole $z$-complex plane with the real branch cut $(x_0,\infty)\subset\R$, with the following three properties:
\begin{enumerate}
\item $f(z)\in\R\,,\forall\, z\in C\cap\R$, i.e., the function is real on the portion of real axis contained in its domain;
\item $f(z)=o(1/\ln(|z|))$ as $z\to\infty$, i.e., the function is an infinitesimal of higher order than $1/\ln(|z|)$ as $z$ diverges;
\item $f(z)\ds\mathop{\propto}_{z\to x_0}(z-x_0)^\sigma$, with $\re(\sigma)>-1$;
\end{enumerate}
the DR for the imaginary part is
\be
f(z)=\frac{1}{\pi}\int_{x_0}^\infty \frac{\im(f(x))}{x-z}dx\,,\hh \forall\, z\in C\,, 
\label{eq:dr-im}
\en
where, as already pointed out, it is understood that the imaginary part is evaluated on the upper edge of the branch cut. If the asymptotic behavior is not the one required at the second item of the previous list, DR's with subtractions have to be used.
\\
More precisely, the DR for the imaginary part subtracted $n$-times at the real point $x_1\in C $, i.e., $x_1<x_0$, is
\be
f(z)\ug
\sum_{k=0}^{n-1}\frac{f^{(k)}(x_1)}{k!}(z-x_1)^k	
	\label{eq:dr-im-subn}\\
	&&+\frac{(z-x_1)^n}{\pi}\int_{x_0}^\infty \frac{\im(f(x))}{(x-x_1)^n(x-z)}dx\,,\hh \forall\, z\in C\,, 
\nen
where $f^{(k)}(z')$ indicates the $k^{\rm th}$ derivative of the function $f(z)$ evaluated at $z=z'$. Such a form of DR is obtained by writing the DR for the imaginary part of Eq.~\eqref{eq:dr-im} for the function
\be
g(z)=\frac{f(z)-\sum_{k=0}^{n-1}\frac{f^{(k)}(x_1)}{k!}(z-x_1)^k}{(z-x_1)^n}\,,
\nen
which has the same analytic properties of the function $f(z)$, with the exception of the asymptotic behavior. Indeed, the limits of the two functions $f(z)$ and $g(z)$, as $z$ diverges, differ by the power $z^n$, i.e., $g(z)=\mathcal{O}\lt {f(z)}/{z^n}\rt$, as $z\to\infty$. It follows that, if the function $f(z)$ had a pole of order $m$ at infinity, so that: $f(z)=\mathcal{O}(z^m)$ as $z\to\infty$, then it would be sufficient to make $n=m+1$ subtractions in order to have $g(z)=\mathcal{O}\lt z^{-1}\rt=o\lt1/\ln(|z|)\rt$. The cost of using the $n$-subtracted DR of Eq.~\eqref{eq:dr-im-subn} is the knowledge of the values of the first $n$ derivatives (from the zero-derivative, the function itself, up to the $(n-1)^{\rm th}$ of the function in the point of subtraction, $z=x_1$.
\\
The simplest form of $n$-subtracted DR is the one with only one subtraction. In this case $n=1$ and the expression of Eq.~\eqref{eq:dr-im-subn} becomes
\be
f(z)\ug f(x_1)
+\frac{z-x_1}{\pi}\int_{x_0}^\infty \frac{\im(f(x))}{(x-x_1)(x-z)}dx\,, 
\label{eq:dr-im-sub1}
\en
$\forall\, z\in C$. The one-subtracted DR has to be used for instance for those functions having a finite, not-null asymptotic value so that: $f(z)=\mathcal{O}(1)$ as $z\to\infty$. In this case, only the value of the function at the subtraction point $z=x_1$ has to be known.
\\
By means of the well-known Sokhotski-Plemelj formula~\cite{plemelj}
\be
\!\!\!\!\!\!\lim _{\e \to 0^{+}}\!\int_{-\infty}^{\infty}
\!{\frac {f(x')}{x'\!-\!x\!-\! i\e }}dx'=\Pr\!\!\int_{-\infty}^{\infty}{\frac {f(x')}{x'\!-\!x}}dx' \!+\! i\pi f(x)\,,
\label{eq:s-p}
\en
where the symbol $\Pr\!\!\int$ stands for the Cauchy principal value integral~\cite{Ahlfors} and the function $f(z)$ is analytic in the infinite rectangle $R=\{z:\im(z)=y\in(\e/2,3\e/2)\}\subset\{z:\im(z)=y>0\}$, belonging to the upper-half $z$-complex plane, and such that
\be
\lim_{x\to\pm\infty}f(x+iy)=0\,,\hh\forall y\in(\e/2,3\e/2)\,,
\nen
 the DR for the imaginary part of Eq.~\eqref{eq:dr-im} can be evaluated also on the upper edge of the real branch cut $(x_0,\infty)$.
\\
In particular, by taking $z=x+i\e$, with $x>x_0$ and $\e\to 0^+$, using the Sokhotski-Plemelj formula of Eq.~\eqref{eq:s-p} and showing explicitly the infinitesimal and positive imaginary part \e, the DR can be written as
\be
f(x+i\e)\ug \frac{1}{\pi}\int_{x_0}^\infty \frac{\im(f(x'+i\e))}{x'-x-i\e}dx'
\no\\
\ug\frac{1}{\pi}\int_{-\infty}^\infty \frac{\im(f(x'+i\e))}{x'-x-i\e}dx'
\no\\
\ug\frac{\Pr}{\pi}\!\!\int_{-\infty}^\infty
\frac{\im(f(x'+i\e))}{x'-x}dx'+i\,\im(f(x+i\e))\,.
\nen
The extension of the integration interval up to the  whole real axis does not imply any additional contribution because the imaginary part of the function $f(z)$ is null for all $z\in(-\infty,x_0]$. By adding $-i\,\im(f(x+i\e))$ to both sides of the previous equation one obtains the so-called DR for the real part, that gives the real part of the function $f(z)$ on the upper edge of the branch cut, i.e.,
\be
\re(f(x))
=\frac{\Pr}{\pi}\!\!\int_{-\infty}^\infty
\frac{\im(f(x'))}{x'-x}dx'\,,
\label{eq:dr-re}
\en
where, as in Eq.~\eqref{eq:dr-im}, once more the infinitesimal imaginary part has to be understood, assuming that both imaginary and real part are always evaluated on the upper edge of the branch cut\footnote{\label{nota:1}Notice that, as a consequence of the Schwarz reflection principle, while the real part of the function assumes the same value on the upper and lower edge of the branch cut, the imaginary part has a discontinuity, by changing the sign, i.e., $\forall\,x>x_0$,
\be
\lim_{\e\to0^+}\re(f(x+i\e))\ug\lim_{\e\to0^+}\re(f(x-i\e))\,,\no
\\
\lim_{\e\to0^+}\im(f(x+i\e))\ug-\lim_{\e\to0^+}\im(f(x-i\e))\,.
\nen}.
\\
Finally, the DR for the real part can be also subtracted,  the expression with $n$ subtractions at $z=x_1<x_0$ follows similarly to the derivation of Eq.~\eqref{eq:dr-im-subn}, in particular, $\forall\, x>x_0$, we have
\be
\re(f(x))\ug
\sum_{k=0}^{n-1}\frac{f^{(k)}(x_1)}{k!}(x-x_1)^k	
	\no\\
	&&+\frac{(x-x_1)^n}{\pi}\Pr\!\!\int_{x_0}^\infty \frac{\im(f(x'))}{(x'-x_1)^n(x'-x)}dx'\,, 
\nen
that, in the simplest case of only one subtraction, becomes
\be
\re(f(x))\ug
f(x_1)\label{eq:dr-re-sub1}\\&&
+\frac{x-x_1}{\pi}\Pr\!\!\int_{x_0}^\infty \frac{\im(f(x'))}{(x'-x_1)(x'-x)}dx'\,. 
\nen
\section{The specific case of $\GE/\GM$}
\label{sec:specific}
The multivalued analytic function under consideration is the FF ratio $R(q^2)=\GE(q^2)/\GM(q^2)$, whose analytic properties are driven from those of the single electric \GE\ and magnetic \GM\ FFs.
\begin{itemize}
\item {\bf Analyticity domain.} Since, as discussed in Sec.~\ref{sec:lambda}, both FFs have the same analyticity domain, that is the $q^2$-complex plane with the real and positive (time-like) branch cut $(\qth,\infty)$, the ratio is a multivalued meromorphic function~\cite{Ahlfors}, having the same branch cut $(\qth,\infty)$ and a set of isolated poles corresponding to the zeros of the magnetic FF \GM, that are not cancelled by equal or higher order zeros of the electric FF \GE. However, we assume that the magnetic FF has no zeros, so that the analytic domain of the ratio does correspond to that of the FFs, i.e., $D_{\rm FF}=\{q^2:q^2\not\in (\qth,\infty)\}$. Moreover, the ratio has zeros in the same points and of the same order as the electric FF \GE.
\\ 
It follows that, having the normalizations at $q^2~=~0$
\be
\begin{aligned}
\GE(0) &= Q_\Lambda= 0\,,\\
\GM(0) &= \mu_\Lambda=-0.613\pm 0.004\,\mu_N\,,
\label{eq:normalizations}
\end{aligned}
\en
where $\mu_N=e\hbar/(2M_pc)$ is the nuclear magneton~\cite{pdg} and, $Q_\Lambda$ and $\mu_\Lambda$ are the electric charge and magnetic moment of the \L\ baryon, the ratio has at least a zero in the origin, as a consequence of the neutrality of the \L\ baryon itself.
\item {\bf Complexity.} The ratio, as well as the FFs, fulfills the Schwarz reflection principle, $R^*(q^2)=R(q^{2*})$, and hence it assumes real values at each real $q^2$ belonging to its analyticity domain, i.e.,
\be
R(q^2)\in\R\,,\hh\forall\,q^2\in D_{\rm FF}\cap\R=(-\infty,\qth)\,.
\nen
On the other hand, it can have a non-null imaginary part at $q_+^2=q^2+i\e$, with $q^2>\qth$ and $\e\to0^+$, that is, on the upper edge of the time-like branch cut. As already discussed in Sec.~\ref{sec:DR}, such an imaginary part represents the source of discontinuity across the branch cut. Indeed, passing through the cut, e.g. from $q_+^2=q^2+i\e$ to $q_-^2=q^2-i\e$ ($q^2>\qth$ and $\e\to0^+$), the imaginary part changes sign, while the real part does not vary being continuous\footnote{See footnote~\ref{nota:1}.},
\be
\lim_{\e\to0^+}\re(R(q^2+i\e))\ug \lim_{\e\to0^+}\re(R(q^2-i\e))\,,
\no\\
\lim_{\e\to0^+}\im(R(q^2+i\e))\ug -\lim_{\e\to0^+}\im(R(q^2-i\e))\,.
\nen
\item {\bf Behavior at the theoretical threshold.} Under the assumption of a magnetic FF having no zeros, at the theoretical threshold $\qth$ the ratio has the same behavior as the electric FF. In particular, following the discussion developed in Sec.~\ref{sec:DR}, at this threshold, corresponding to the branch point of the cut, the electric and magnetic FFs behave like the powers 
\be
G_{E,M}^\Lambda(q^2)\mathop{\propto}_{q^2\to\qth} (q^2-\qth)^{\gamma_{E,M}}\,,
\nen
with $\re(\gamma_{E,M})>-1$. Moreover, since the magnetic FF has no zeros, the real part of the power $\gamma_M$ can not be greater or equal to one, so that it must fulfill the condition: $-1<\re(\gamma_M)\le 0$. Therefore, the behavior of the ratio at the theoretical threshold fulfills the requirement of the DR's, i.e.,
\be
R(q^2)\mathop{\propto}_{q^2\to\qth} (q^2-\qth)^{\gamma_{E}-\gamma_{M}}\,,
\nen
with $\re(\gamma_{E}-\gamma_{M})>-1$, as follows from the limitations: $\re(\gamma_E)~>~-1$ and $-\re(\gamma_M)\ge0$.
\item {\bf Asymptotic behavior.} The high-$|q^2|$ behavior of the FFs is obtained in the framework of perturbative QCD, that predicts the same power-law for both FFs~\cite{Matveev:1973uz,Brodsky:1973kr}, namely
\be
G^{\Lambda}_{E,M}(q^2)=\mathcal{O}\lt(q^2)^{-2}\rt\,, \ \ \ q^2\to -\infty\,.
\label{eq:asy}
\en
In fact, at high $q^2$ the dominant amplitude of the space-like process $\gamma^*B\to B$, where the virtual photon $\gamma^*$ is absorbed by the baryon $B$, is that containing two gluon propagators and hence it scales like $(q^2)^{-2}$. This is a consequence of the fact that, in order to maintain intact the baryon, the four-momentum transferred by the virtual photon has to be shared among the constituent quarks through multiple gluon exchanges and, at high $q^2$, the most probable reaction mechanism is that involving the minimum number of gluon exchanges, that are needed for the complete four-momentum distribution among the three constituent quarks of the baryon.
\\
 Moreover, the analyticity implies that the FFs have to have the same asymptotic behavior in space-like and time-like regions, and hence the high-$(-q^2)$ trend described in Eq.~\eqref{eq:asy} is valid also in the time-like limit $q^2\to\infty$. 
\\
It is straightforward to infer that the ratio of FFs is asymptotically constant. This is the obvious consequence of the fact that the high-$|q^2|$ trend of both FFs is ruled by the same power-law, which is given in Eq.~\eqref{eq:asy}. Hence, by neglecting QCD corrections, the ratio scales like
\be
R(q^2)=\mathcal{O}(1)\,,\hh |q^2|\to\infty\,.
\nen
On the other hand, the inclusion of QCD corrections does affect the high-$|q^2|$ behavior of the ratio. However, there is no consensus on how to implement such corrections, different methods lead to different results for the asymptotic behavior of FFs and $R(q^2)$. Two examples are described in App.~\ref{app:qcd-corr}.\\
Even though in some scheme QCD-corrections provide a logarithmic divergent trend for the ratio $R(q^2)$ as $|q^2|\to\infty$, DR's with a single subtraction, see Sec.~\ref{sec:DR}, remain still applicable, because the convergence condition $R(q^2)/q^2=o\lt\ln(|q^2|)\rt$, as $|q^2|\to\infty$, is verified.
\\
Finally, from the extension of the power-law behavior of Eq.~\eqref{eq:asy} also to the time-like limit $q^2\to\infty$, it follows that 
\be
\lim_{q^2\to \infty} \frac{G_{E,M}^\Lambda(q^2)}{G_{E,M}^\Lambda(-q^2)}=1\,.
\label{eq:sl-tl-lim}
\en
This means that not only FFs scale with the same power-law but also that they have just the same time-like and space-like limits. Thus their ratio, time-like over space-like in Eq.~\eqref{eq:sl-tl-lim}, tends to unity. A further consequence is that, since the time-like FFs are complex functions with non-null imaginary parts, such imaginary parts have to vanish faster than the real ones in the limit $q^2\to\infty$.
\end{itemize}
\section{The parameterization for $R(q^2)$ and the $\chi^2$ definition}
\label{sec:parameterization}
\begin{figure}[H]
\begin{center}
		\includegraphics[width=.9\columnwidth]{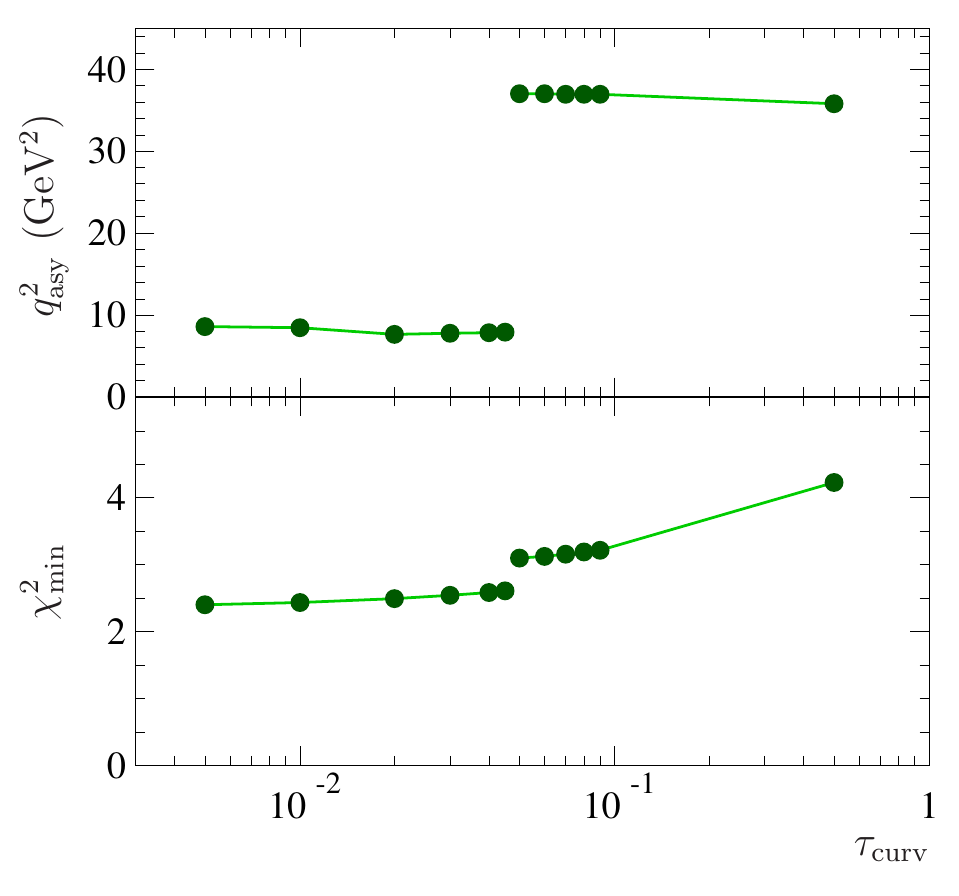}\end{center}\vspace{-8mm}
		\caption{\label{fig:tau-chi2} Best values of the parameter \qasy\ and minima of the total $\chi^2$ defined in Eq.~\eqref{eq:chi2} as a function of the regularization parameter $\tau_{\rm curv}$, with $N=5$.}
\end{figure}

The imaginary part of the ratio is parametrized as a combination of Chebyshev polynomials of the first kind $T_j(x)$~\cite{abramovitz}, with $j\in\N\cup\{0\}$ and $x\in[-1,1]$. These polynomials can be defined through the trigonometric form
\be
T_j\lt\cos(\alpha)\rt=\cos(j\alpha)\,,
\nen
with $\alpha\in[0,2\pi]$. In this form, the values at the extremes $x=+1,-1$, corresponding to $\alpha=0,\pi$, are
\be
T_j(-1)= \cos(j\pi)=(-1)^j\,,\hh
T_j(1)=	\cos(0)=1\,.
\nen
We define the parameterization 
\be
\im \lq R(q^2)\rq &\equiv& Y\big( q^2;\vec{C},\qasy\big) 
\label{eq:cheby}\\
Y\big( q^2;\vec{C},\qasy\big) \ug
\left\{
\begin{array}{ll}
	\ds\sum_{j=0}^{N}C_jT_j\lq x (q^2)\rq &  \qth<q^2<\qasy\\
&\\
0 & q^2\ge \qasy\\
\end{array}
\right.,
\nen
where 
\be
x(q^2)=2\frac{q^2-\qth}{\qasy-\qth}-1\,.
\nen
It depends on the set of $N+2$ free parameters $\{\vec C=(C_0,C_1,\ldots,C_N),\qasy\}$, namely $N+1$ coefficients of the Chebyshev polynomials, representing the components of the vector $\vec C\in\R^{N+1}$, together with the asymptotic threshold $\qasy\in\big( \qphy,\infty\big)$.
\\
The free parameters are determined by imposing theoretical and experimental constraints. While for the latter a standard $\chi^2$ minimization procedure is adopted, the former, which actually represent a set of conditions that have to be exactly fulfilled, are forced by means of two different methods. Those regarding directly the imaginary part, i.e., known values at some well established $q^2$, are forced by fixing a subset of the Chebyshev coefficients. The other theoretical constraints, which instead concern the real part, obtained from the imaginary part through the subtracted DR of Eq.~\eqref{eq:dr-re-sub1}, are imposed by means of a $\chi^2$ minimization. 
\\
The theoretical constraints directly on the imaginary part,  parametrized as given in Eq.~\eqref{eq:cheby}, follow from the reality of the FF ratio at the theoretical and physical thresholds, as well as in the asymptotic region $\big(\qasy,\infty\big)$. They can be imposed by the following three conditions 
\be
\left\{\begin{array}{lcl}
R(\qth)\in\R & \Rightarrow & Y\big(\qth;\vec C,\qasy\big)=0\\	&&\\
R(\qphy)\in\R & \Rightarrow & Y\big(\qphy;\vec C,\qasy\big)=0\\
	&&\\
R(q^2\ge\qasy)\in\R & \Rightarrow & Y\big(q^2\ge\qasy;\vec C,\qasy\big)=0\\
\end{array}
\right..\;\;\;\;\;\;\;
\label{eq:nodes}
\en
These identities 
represent three equations of a linear system having as unknowns three arbitrary Chebyshev coefficients, e.g., $C_0$, $C_1$ and $C_2$, so that the number of degrees of freedom for the parametrization of Eq.~\eqref{eq:cheby} reduces from $N+2$ to $N-1$, i.e., $N-2$ Chebyshev coefficients together with the asymptotic threshold \qasy. 
\\
The other theoretical constraints concern the value of the real part of the ratio. In particular, at the physical threshold \qphy, where the FFs are equal each other, the ratio is not only real, but is equal to one and hence the condition is 
\be
\re\big[ R(\qphy)\big]=\frac{\qphy}{\pi}\Pr\!\int_{\qth}^{\qasy}\frac{Y\big( s;\vec C,\qasy\big)}{s(s-\qphy)}ds=1\,.
\nen
\\
It is also considered the possibility of forcing the convergence to the unity, as $q^2\to\infty$, of the modulus of real part $\re\big[ R(q^2)\big]$, in accordance with the perturbative QCD prediction. It is assumed that such a limit value is attained at the asymptotic threshold \qasy, and constantly maintained for $q^2\ge\qasy$. This eventuality is accounted for by requiring
\be
\left|\re\lq R(q^2)\rq\right|=\frac{q^2}{\pi}\Bigg|\Pr\!\int_{\qth}^{\qasy}\frac{Y\big( s;\vec C,\qasy\big)}{s(s-q^2)}ds\Bigg|=1\,,
\nen
for $q^2\ge\qasy$. \\
Concerning the experimental conditions, as already discussed, the only available data are:
\begin{itemize}
\item on the modulus of the ratio, $|R(q^2)|$: two points from BESIII and one point from BaBar, the corresponding  set is $\{q^2_j,|R_j|,\delta|R_j|\}_{j=1}^M$, with $M=3$;
\item on the sinus of the phase $\phi(q^2)=\arg\big(R(q^2)\big)$: one point from BESIII and one from BaBar, the set is $\{q^2_k,\sin(\phi_k),\delta\sin(\phi_k)\}_{k=1}^P$, with $P=2$.
\end{itemize}
The $\chi^2$, depending on the set of $N-1$ free parameters $\{C_3,C_4,\ldots,C_N,\qasy\}$ (the first three Chebyshev coefficients have been fixed by the linear system of Eq.~\eqref{eq:nodes}), and accounting for all these conditions, is defined as
\be
\chi^2\big(\vec C,\qasy\big)\ug
\chi^2_{|R|}+\chi^2_{\phi}
\no\\&&+\tau_{\rm phy}\chi^2_{\rm phy}
+\tau_{\rm asy}\chi^2_{\rm asy}
+\tau_{\rm curv}\chi^2_{\rm curv}\,.\hh\hh
\label{eq:chi2}
\en  
The first two contributions, imposing the experimental constraints, are
\be
\chi^2_{|R|}\ug \sum_{j=1}^M\lt\frac{ \sqrt{X\big(q^2_j
\big)^2-Y\big(q^2_j
\big)^2}-|R_j|}{\delta |R_j|}\rt^2\,,\no\\
\chi^2_{\phi}\ug \sum_{k=1}^P\lt\frac{ \sin\lt\arctan\lt {Y\big(q^2_k
\big)}/{X\big(q^2_k
\big)}\rt\rt-\sin(\phi_k)}{\delta\sin(\phi_k)}\rt^2\,,
\nen
where $X(q^2)$ is the real part obtained through the DR
\be
X(q^2)\equiv\re\big(R(q^2)\big)=\frac{q^2}{\pi}\Pr\!\int_{\qth}^{\qasy}\frac{Y(s)}{s(s-q^2)}ds\,.\hh
\label{eq:re}
\en
The third and fourth contributions impose, with adjustable weights $\tau_{\rm phy}$ and $\tau_{\rm asy}$, the constraints on the real part of the ratio at the physical and asymptotic thresholds, \qphy\ and \qasy, i.e.,
\be
\chi^2_{\rm phy}= \lt 1-X(\qphy)\rt^2\,,
\hh
\chi^2_{\rm asy}= \lt 1-X(\qasy)^2\rt^2\,.
\nen 
The latter is written for the squared value, because the perturbative QCD prediction concerns the modulus of the real part and hence both values $X(\qasy)=\pm 1$ have to be considered. 
\\
The last contribution $\tau_{\rm curv}\chi^2_{\rm curv}$ is introduced to stabilize the solution of a so-called ill-posed problem, usually represented by a Fredholm integral equation~\cite{fredholm}.
\\ 
For instance, the inhomogeneous Fredholm integral equation of the second kind 
\be
u(x)=\int_{c}^d K(x,x')u(x')dx'+\phi(x)\,,\hh
\begin{array}{r}
x\in[a,b]\\
x'\in[c,d]\\
\end{array}\,,
\nen
where $\phi(x)$, $u(x)$ and $K(x,x')$ are the input function, the unknown function and the kernel, can be solved by using the Tikhonov regularization procedure~\cite{fredholm}. Such a procedure consists in minimizing the functional 
\be
F[u]\ug \int_{a}^b\bigg| u(x)-\phi(x)-\int_c^d K(x,x')u(x')dx'\bigg|^2 dx\no\\
&&+\tau\, C[u]\,,
\nen
where the interval $[a,b]\subset\R$ is the domain of the input function $\phi(x)$, the real non-negative number $\tau$ is the  regularization parameter and the functional $C[u]$ represents the total curvature of the solution  $u(x)$ for $x\in[c,d]$, i.e., 
\be
C[u]=\int_c^d\bigg|\frac{d^2u}{dx^2}\bigg|^2dx\,.
\nen
It follows that the total curvature of the solution obtained by minimizing the functional $F[u]$ with a certain value of $\tau$, is large for a small value of $\tau$, and small for a large value of $\tau$. The extreme cases are those with $\tau\to\infty$, giving constant $u(x)$, and $\tau=0$ resulting, instead, in a highly-oscillating solution.
\\
In our case the functional to be minimized is the $\chi^2$ of Eq.~\eqref{eq:chi2} and the unknown function is $Y(q^2)$, the imaginary part of the ratio defined in Eq.~\eqref{eq:cheby}. The real part $X(q^2)$ is given by the DR integral of Eq.~\eqref{eq:re}, and it is just this quantity that gives the "integral" character to the minimization problem. As a consequence, the contribution $\chi^2_{\rm curv}$ to the total $\chi^2$ has the form
\be
\chi^2_{\rm curv}=\int_{\qth}^{\qasy}\bigg|\frac{d^2Y(s)}{ds^2}\bigg|^2 ds\,,
\nen
indeed, it does represent the total curvature of $Y(q^2)$ in the $q^2$ interval $[\qth,\qasy]$ where it is not constant. 
\\
The values of the weights $\tau_{\rm phy}$ and $\tau_{\rm asy}$ have been chosen in order to make the fulfillment exact. In particular, it happens that starting from a threshold value, namely $\tau_{\rm phy,asy}=\tau_{\rm phy,asy}^0$, the corresponding $\chi^2$ contribution turns out to be negligible with respect to the others, so that the minimization procedure becomes independent of their values. On the basis of these considerations, the minimization procedure has been carried out by setting $\tau_{\rm phy}=\tau_{\rm phy}^0$, and $\tau_{\rm asy}=\tau_{\rm asy}^0$.
\\
On the contrary, the value of the regularization parameter $\tau_{\rm curv}$, that, through the $\chi^2$ minimization procedure, allows to control the total curvature of $Y(q^2)$ in the $q^2$ interval $[\qth,\qasy]$, has to be determined by means of a more refined technique.
\\
Indeed, the parameterization in terms of the combination of the first $N+1$ Chebyshev's polynomials, and hence the degree $N$ of the resulting polynomial in $q^2$, given in Eq.~\eqref{eq:cheby}, represents by itself a solution with a controlled curvature. No more than $N$ extrema, maxima plus minima, i.e., zeros of the first derivative of the the polynomial, can be present in any solution. This means that the number of oscillations is limited. On the other hand, the constraint represented by the DRs integral limits the amplitude of such oscillations. It follows that the combined effect of these two features of the parametrization does not allow largely unstable solutions. Nevertheless, the ill-posedness nature of the inverse problem has to be faced, because, when the regularization parameter $\tau_{\rm curv}$ is set to zero, the solution always has the maximum curvature, i.e., the maximum number of oscillations allowed by the degree of the polynomial. 
\\
In the light of that, the degree $N$ and the regularization parameter $\tau_{\rm curv}$ are mutually dependent. The selection of their values has to be made by suitably balancing two opposite tendencies: the increase of the total curvature as the polynomial degree increases; the progressive suppression of the oscillations, and hence the reduction of the curvature, achieved by increasing the value of the regularization parameter $\tau_{\rm curv}$. 
\\
In addition, to select the final values of $N$ and $\tau_{\rm curv}$, we also consider the role played by the free parameter \qasy, representing the $q^2$ threshold from which the asymptotic behavior is assumed. In particular, solutions with larger values of such a free parameter are evidently privileged. 
\\
Figure~\ref{fig:tau-chi2} shows the best values of \qasy\ and the corresponding $\chi^2$ minima with $N=5$ as a function of $\tau_{\rm curv}$, at which the minimization is performed. At $\tau_{\rm curv}=0.05$ there is a kind of phase transition where the best value for the asymptotic threshold moves from $\qasy\simeq8$~GeV$^2$ to a larger value $\qasy\simeq37$~GeV$^2$, which is more reliable according to its own definition. As a consequence of such a widening of the $q^2$ interval covered by the polynomial parametrization, namely $[\qth,\qasy]$, there is a limited but acceptable ($\le20\%$) increase of the minimum $\chi^2$. The final values for the regularization parameter and the highest polynomial degree have then  been chosen as \be
\tau_{\rm curv}=0.05\,,\hh N=5\,.
\label{eq:tau_curv-N}
\en
\noindent
\section{Results and discussion}
\label{sec:results}
\begin{figure}[H]
\begin{center}
\includegraphics[width=.9\columnwidth]{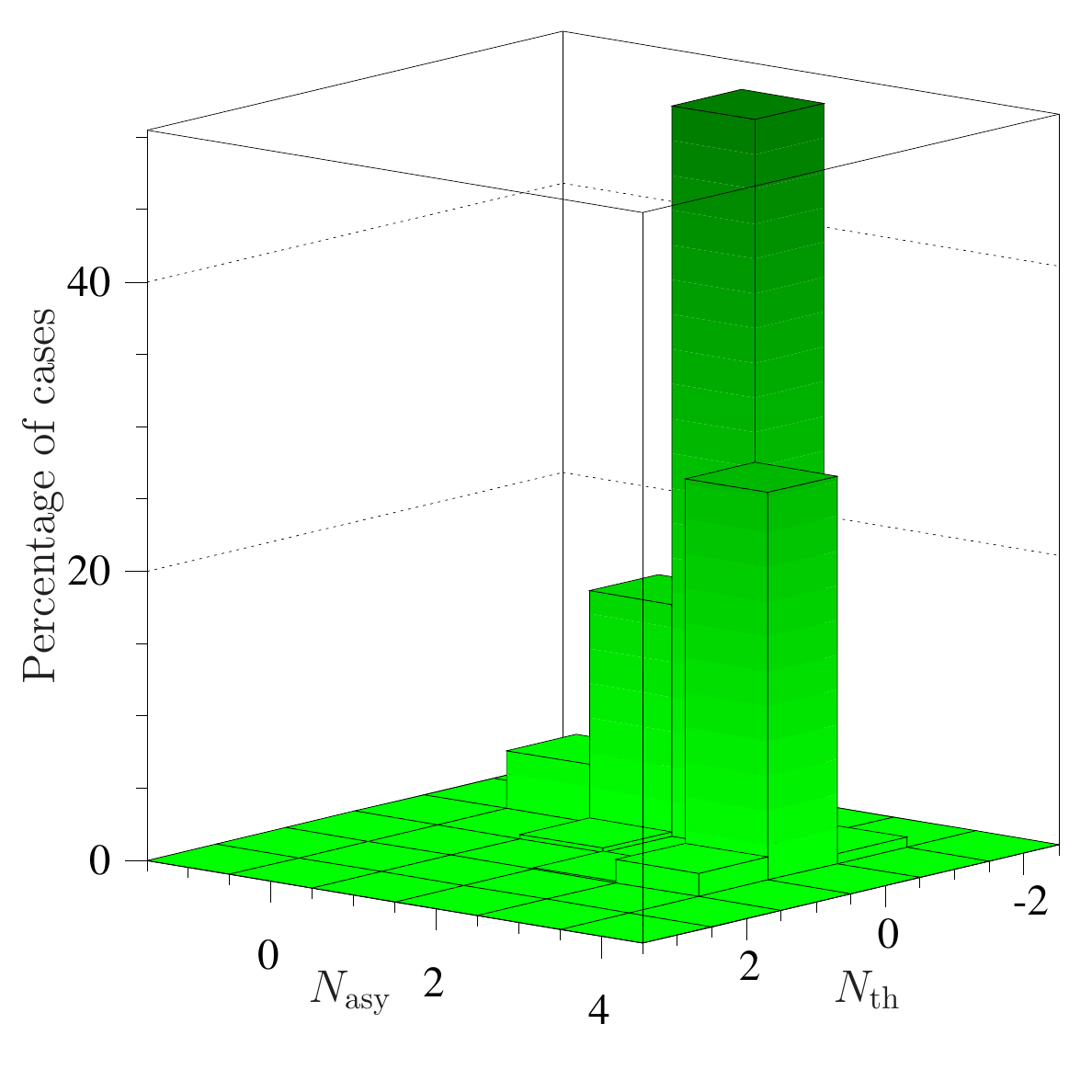}	
\end{center}\vspace{-8mm}
\caption{\label{fig:cases}Percentage of cases characterized by the values of the phases, in units of $\pi$ radiants, at the theoretical threshold \qth\ and at \qasy.}
\end{figure}

\begin{figure}[H]	
\begin{center}
\includegraphics[width=.49\columnwidth]{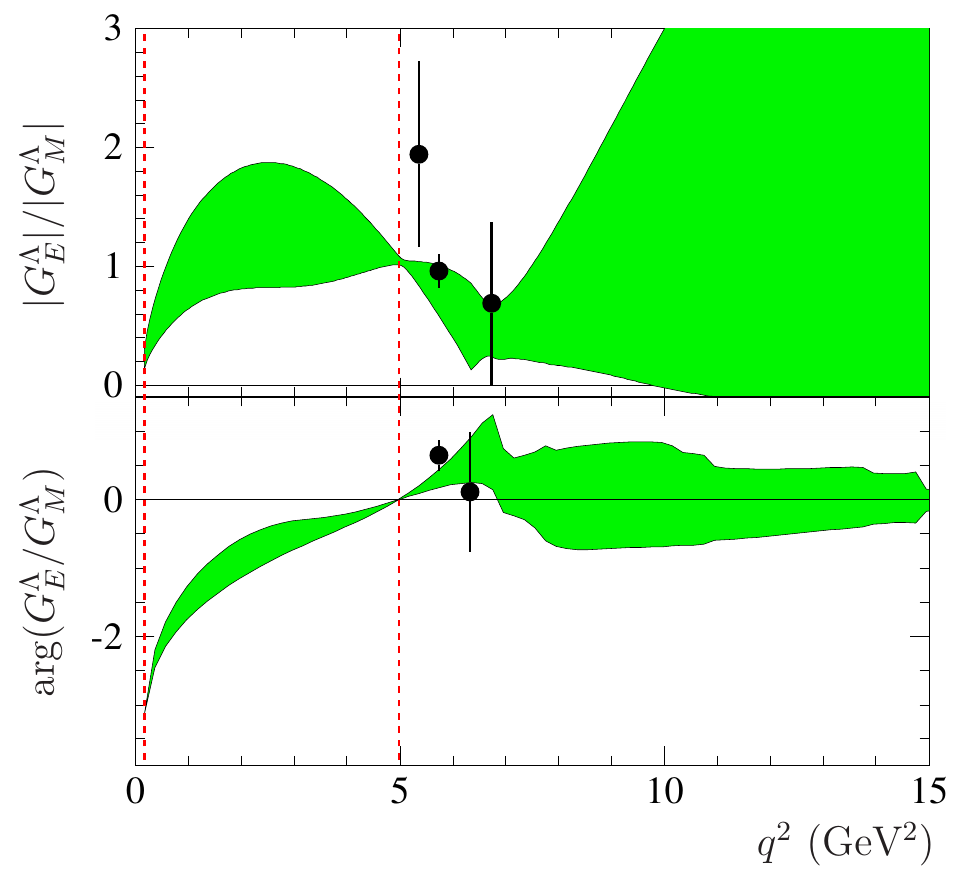}	
\includegraphics[width=.49\columnwidth]{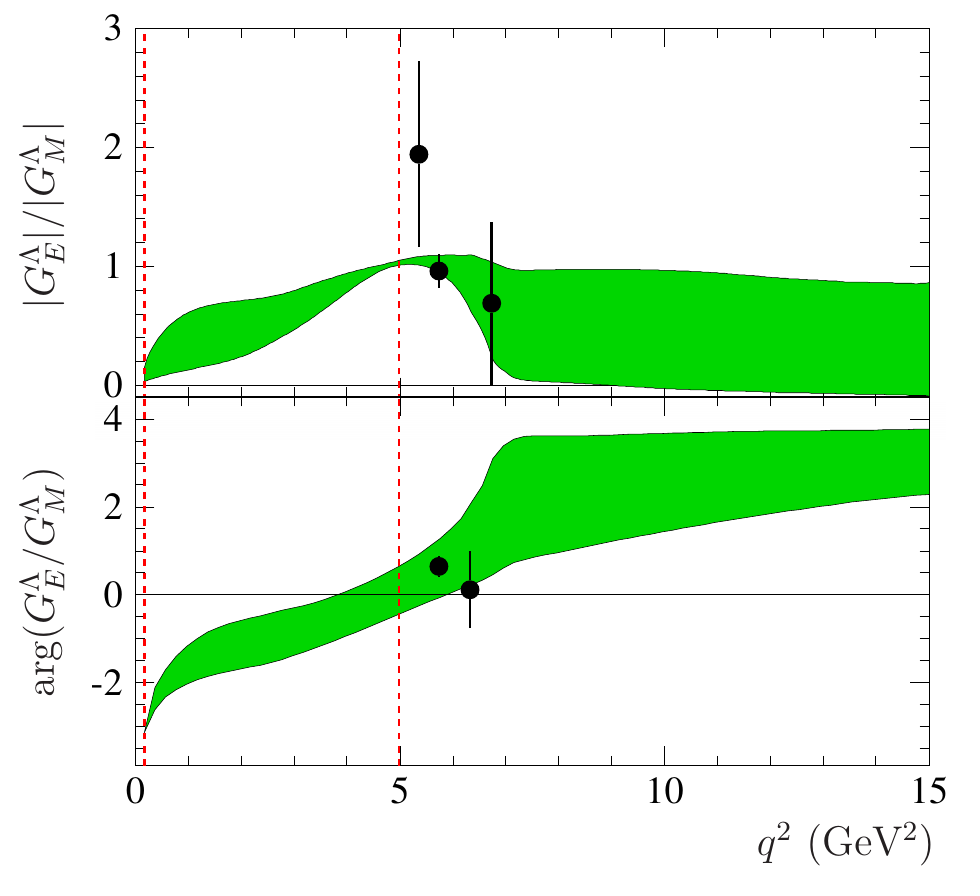}	
\includegraphics[width=.49\columnwidth]{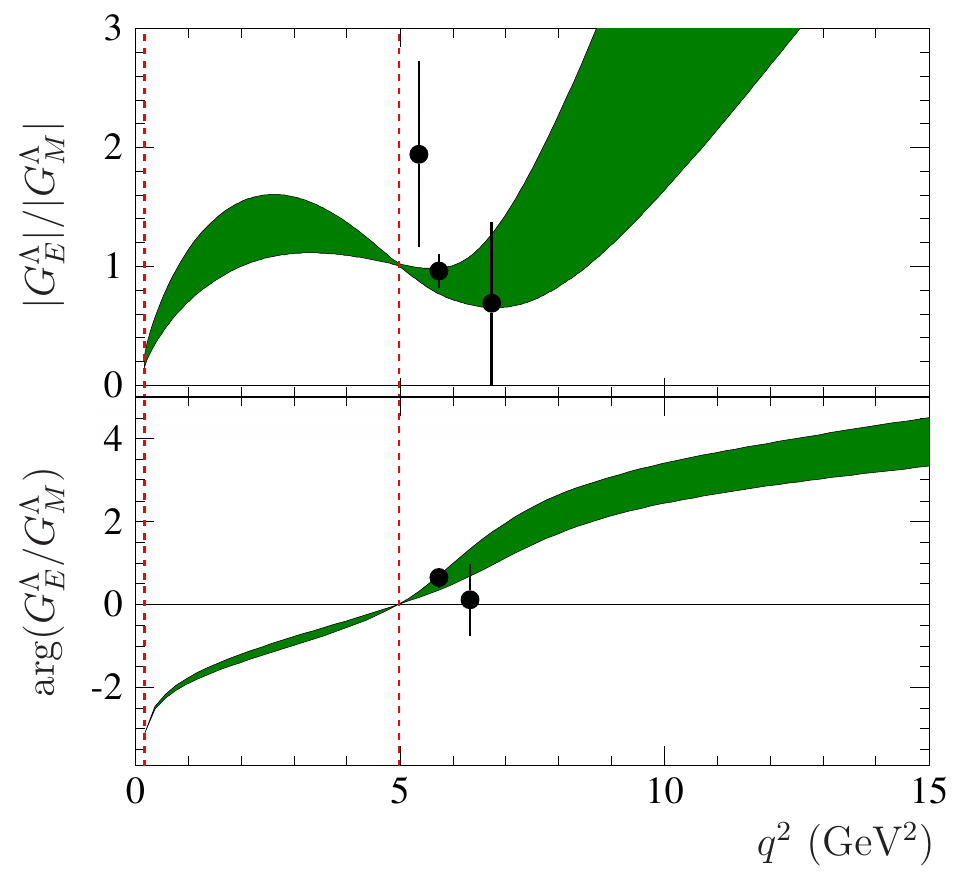}	
\includegraphics[width=.49\columnwidth]{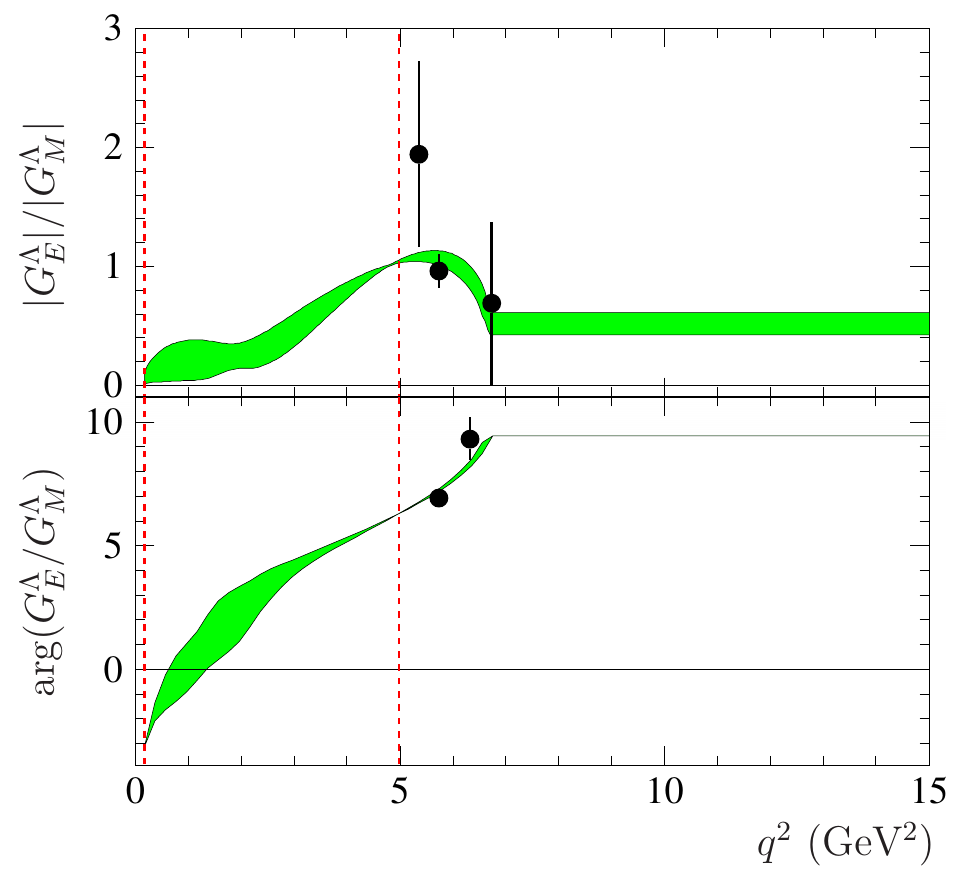}	
\includegraphics[width=.49\columnwidth]{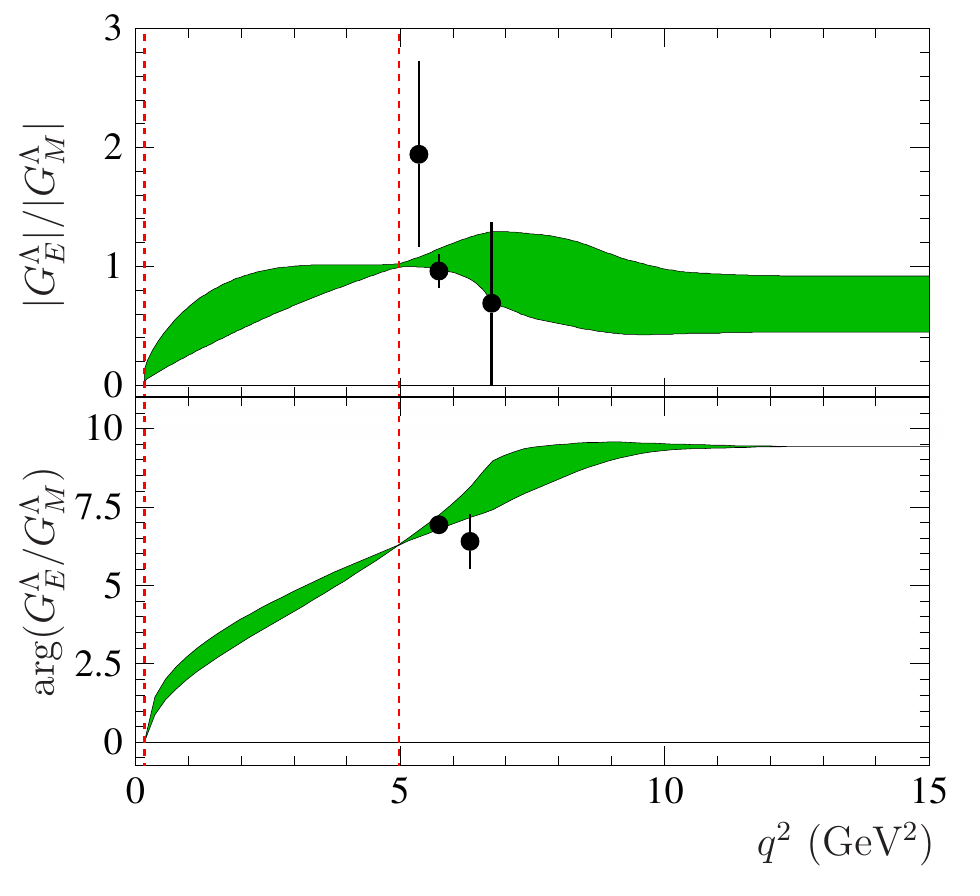}	
\includegraphics[width=.49\columnwidth]{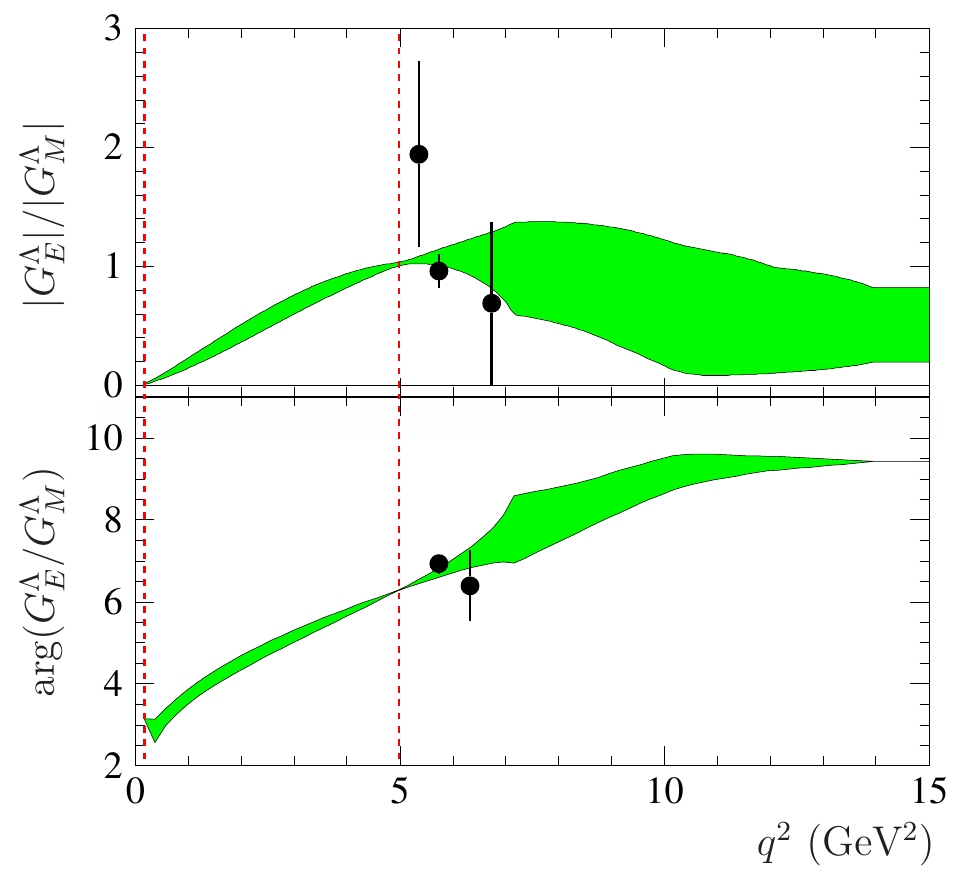}	
\caption{\label{fig:moduli-phases}Modulus (upper panels) and phase (lower panels) of the FF ratio with: $(\nth,\nasy)=(-1,0)$ in position left-up, $(\nth,\nasy)=(-1,1)$ in position right-up, $(\nth,\nasy)=(-1,2)$ in position left-middle, $(\nth,\nasy)=(-1,3)$ in position right-middle, $(\nth,\nasy)=(0,3)$ in position left-down, $(\nth,\nasy)=(1,3)$ in position right-down. The color intensity of the error bands is proportional to the occurrence probability of the corresponding pair $(\nth,\nasy)$, a darker color indicates an higher probability. The red dashed lines indicate the theoretical and the physical thresholds.
}\end{center}
\end{figure}

The very few available data and their large uncertainties represent an obstacle to the complete determination of the FF ratio as a complex function of $q^2$ in the time-like region, above the theoretical threshold $\qth$ and, as a real function of $q^2$ in the remaining kinematical range. Since, as already stated, at these energies the FFs and hence their ratio are real, the relative phase assumes values that are integer multiples of $\pi$ radiants. Such an eventuality is accounted for by defining the corresponding integers $\nth$ and $\nasy$ as
\be	
\nth=\frac{1}{\pi}\arg\lt\!\frac{\GE(\qth)}{\GM(\qth)}\!\rt,\ 
\nasy= \frac{1}{\pi}\arg\lt\!\frac{\GE(\qasy)}{\GM(\qasy)}\!\rt.
\nen
As already widely discussed and exhaustively argued in the previous sections, the theoretical constraints based on the first principles of Quantum Field Theory and analyticity of the FFs offer a unique possibility to achieve educated guesses on the phase determination, by relating through the DR of Eq.~\eqref{eq:re}, imaginary and real part and hence, modulus and phase of the ratio $\GE/\GM$. It is important to reiterate that this phase determination embodies crucial physical information which is not experimentally accessible, since 
the experiments measure only the sinus of the phase.
\\
Unfortunately, the only available five data points with their large uncertainties do not allow to obtain unique values for \nth\ and \nasy. Nevertheless, the number of possible solutions, that in principle are infinite, is drastically reduced to the few possibilities reported in Table~\ref{tab:cases}. Furthermore, the Monte Carlo procedure,  built to perform a statistical analysis of our results, gives also the probability of occurrence of each pair $\lt\nth,\nasy\rt$. These probabilities are reported as percentages and visual strips in Table~\ref{tab:cases} and as a three-dimensional histogram on a $\nth\nasy$-grid in Fig.~\ref{fig:cases}.
\\
\begin{table}[h!]
\begin{center}
\begin{tabular}{c|c|r| l}
 \nth & \nasy & \% & Visual percentage\\
 \hline
\hline
-1 & 0 & 4.0 & {\color{verde1}\rule{0.02667\textwidth}{2 mm}}\\ \hline
-1 & 1 & 16.0 & {\color{verde2}\rule{0.1067\textwidth}{2 mm}}\\ \hline
-1 & 2 & 50.5 & {\color{verde3}\rule{.3367\textwidth}{2 mm}}\\ \hline
-1 & 3 & 0.7 & {\color{verde4}\rule{0.0047\textwidth}{2 mm}}\\ \hline
0 & 1 & 0.3 & \rule{.0020\textwidth}{2 mm} \\ \hline
0 & 3 & 26.8 & {\color{verde5}\rule{.1787\textwidth}{2 mm}}\\ \hline
1 & 2 & 0.1 & \rule{.00067\textwidth}{2 mm}\\ \hline
1 & 3 & 1.6  & {\color{verde6}\rule{0.01067\textwidth}{2 mm}}\\ \hline
\end{tabular}		
\caption{\label{tab:cases}Numerical and visual percentage of cases corresponding to the values of the phase, in units of $\pi$ radiants, at the theoretical threshold \qth\ and at \qasy.}
\end{center}
\end{table}
\noindent
We classify in terms of the pairs $\lt\nth,\nasy\rt$ the different descriptions of the two observables, phase and modulus of the FF ratio $\GE/\GM$, obtained through our procedure, which is based on DRs and experimental data.
\\
By considering as acceptable the cases having a probability of occurrence larger than 0.5\%, only six pairs $\lt\nth,\nasy\rt$ are phenomenologically compatible with the data. Once again, the fact that so few cases are selected despite the lack of experimental knowledge of the pivotal observables, namely the modulus and the phase of the ratio $G_E^\Lambda/G_M^\Lambda$, proves the strength and the effectiveness of the theoretical frame of the procedure. 
\\
The results obtained for the modulus and the phase of $G_E^\Lambda/G_M^\Lambda$ corresponding to the six cases $\lt\nth,\nasy\rt=(-1,0),\,(-1,1),\,(-1,2),\,(-1,3),\,(0,3),\,(1,3)$ are shown in Fig.~\ref{fig:moduli-phases}, from the left-up to the right-down panel respectively.
\\
At the theoretical threshold the phase of the ratio can assume only three possible values: $0$ or $\pm\pi$, i.e., $\nth=0$ or $\nth=\pm1$. They depend on the sign of the real values of the FFs \GE\ and \GM\ just below the theoretical threshold \qth, above which they acquire a non-vanishing imaginary part. In particular:
\begin{itemize}	
\item $\nth=0$, if the FFs \GE\ and \GM\ at $q^2\to{\qth}^-$ have the same sign, i.e.,
\be\left\{\begin{array}{rcl}	
 \GE(q^2) & \ds\mathop{\longrightarrow}_{q^2\to{\qth}^-}&   +(-)|\GE(\qth)|=|\GE(\qth)|e^{0(i\pi)}
\\
\GM(q^2) & \ds\mathop{\longrightarrow}_{q^2\to{\qth}^-}& +(-)|\GM(\qth)|=|\GM(\qth)|e^{0(i\pi)}\\
\end{array}\right.\,;
\nen
\item $\nth=+1$, if at $q^2\to{\qth}^-$ the electric FF is negative and the magnetic is positive, 
\be\left\{\begin{array}{rcl}	
 \GE(q^2) & \ds\mathop{\longrightarrow}_{q^2\to{\qth}^-}&  -|\GE(\qth)|=|\GE(\qth)|e^{i\pi}
\\
\GM(q^2) & \ds\mathop{\longrightarrow}_{q^2\to{\qth}^-}& |\GM(\qth)|=|\GM(\qth)|e^0\\
\end{array}\right.\,;
\nen
\item $\nth=-1$, if at $q^2\to{\qth}^-$ the electric FF is positive and the magnetic is negative, 
\be\left\{\begin{array}{rcl}	
 \GE(q^2) & \ds\mathop{\longrightarrow}_{q^2\to{\qth}^-}&  +|\GE(\qth)|=|\GE(\qth)|e^0
\\
\GM(q^2) & \ds\mathop{\longrightarrow}_{q^2\to{\qth}^-}& -|\GM(\qth)|=|\GM(\qth)|e^{i\pi}\\
\end{array}\right.\,.
\nen
\end{itemize}
Concerning the asymptotic behavior, we assume that from $q^2=\qasy$ up to infinity the FF ratio stays real, and hence the phase in this $q^2$ region is constant and equal to the integer multiple of $\pi$ radians $\nasy\pi$. Moreover, assuming that the magnetic FF has no zeros in the $q^2$-complex plane with the branch cut $(\qth,\infty)$, so that the ratio, having no poles, is analytic in the same domain, we can invoke the Levinson's theorem, Eq.~\eqref{eq:levinson}, to infer that \nasy\ must be greater than \nth. In the peculiar case of the neutral $\Lambda$ baryons, the normalization at the origin of the electric FF to the total charge, i.e., $\GE(0)=0$, does imply
\be
\nasy\ge \nth+1\,.
\nen 
The unity in the right-hand side accounts for the zero at $q^2=0$ of the ratio $\GE/\GM$, which, indeed, is the consequence of the zero of \GE\ and of the not vanishing value of  $\GM(0)=\mu_\Lambda\not=0$, as it is shown in Eq.~\eqref{eq:normalizations}.
\noindent
Figure~\ref{fig:moduli-phases} is a grid of six double graphs, which show the modulus and the phase of the ratio in the upper and lower panel, respectively, for each of the six pairs of $(\nth,\nasy)$ having an occurrence probability larger than 0.5\%, see Table~\ref{tab:cases}. It is interesting to notice how the procedure works in establishing the determination of the phase. Indeed, different determinations are assigned to the phase values extracted from the same data on its sinus (insensitive to the determination), once that the connection with the modulus is established through DRs and a set of crucial theoretical constraints is imposed. In particular, the phase values extracted from the most precise datum on the sinus, namely that provided by the BESIII Collaboration~\cite{Ablikim:2019vaj}, shown as dark green square in the right panel of Fig.~\ref{fig:data-LL}, has been used as the reference point to which is assigned the fundamental determination, namely $2k\pi<\arg(\GE/\GM)<(2k+1/2)\pi$, being $0<\sin\lt\arg(\GE/\GM)\rt<1$, and with the integer $k\in\N\cup\{0\}$ obtained by the dispersive analysis.
\\
The determination of the BaBar point has been fixed as a consequence of this choice and on the light of the result of the dispersive analysis, which selected one of the two possibilities, either $2k\pi<\arg(\GE/\GM)<(2k+1/2)\pi$, as for the BESIII point, or $(2k+1/2)\pi<\arg(\GE/\GM)<(2k+1)\pi$. As can be seen in the lower panel of the right-middle picture of Fig.~\ref{fig:moduli-phases}, only in the case with $(\nth,\nasy)=(-1,3)$, where $k=1$, the two fundamental determinations are different, in fact, the BESIII phase lies in $2\pi<\arg(\GE/\GM)<5\pi/2$, while the BaBar value in $5\pi/2<\arg(\GE/\GM)<3\pi$. More in detail, the value of $k$ is zero in the first three cases, with $(\nth,\nasy)=(-1,0),\,(-1,1),\,(-1,2)$, while it is $k=1$ in the last three cases: $(\nth,\nasy)=(-1,3),\,(0,3),\,(1,3)$.
\begin{figure}[htb]	
\begin{center}
\includegraphics[width=\columnwidth]{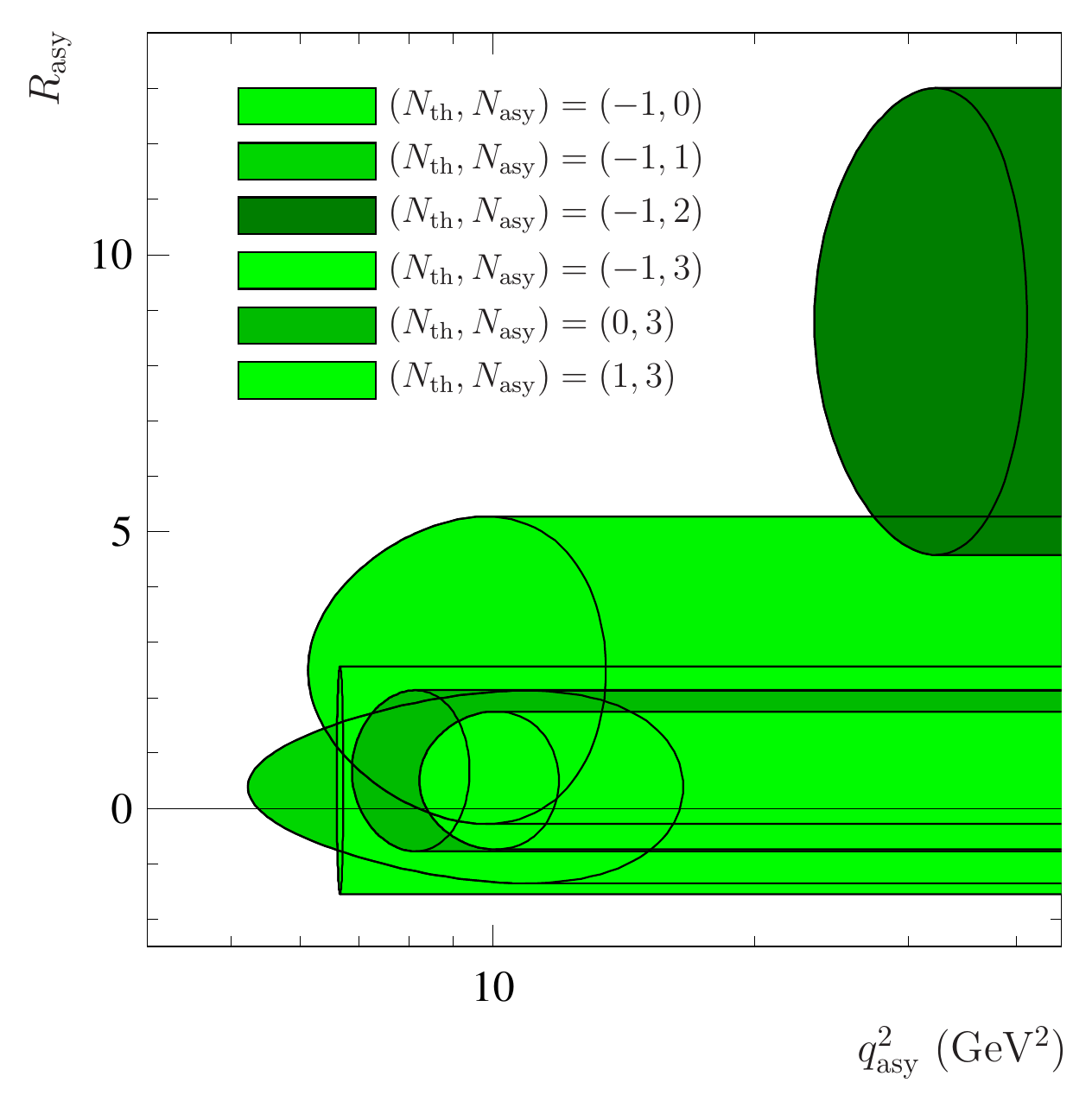}
\caption{\label{fig:sasys}Asymptotic behaviors for each case $(\nth,\nasy)$. The ellipses (deformed by the logarithmic scale) are centered at $((\qasy,\Rasy))$, with $\Rasy\equiv |R(\qasy)|$, and have the errors $\delta\qasy$ and $\delta\Rasy$ as horizontal and vertical semi-axes, respectively. The right-hand side tails represent the error bands of the asymptotic constant values.
}\end{center}
\end{figure}\noindent
\subsection{The modulus of $\GE/\GM$}
\label{subsec:modulus}
The modulus of the FF ratio represents the best-established observable of this analysis since it is not affected by any ambiguity of definition as it does, instead, the phase. Nevertheless, the lack of data, makes the predictions quite uncertain especially for the asymptotic behavior, which, in the majority of the cases, concerns a $q^2$-region lying well above the higher available data point. 
\\
In particular, the procedure gives as outcomes for each pairs $(\nth,\nasy)$ the asymptotic threshold $\lt\qasy\pm\delta\qasy\rt$, as well as the corresponding values of the modulus of the ratio $\lt \Rasy\pm \delta \Rasy\rt$, where $\Rasy$ is defined as $\Rasy\equiv |R(\qasy)|$. The obtained values of these asymptotic parameters are reported in Fig.~\ref{fig:sasys} as ellipses centered at $(\qasy,\Rasy)$ and having as horizontal and vertical semi-axes the uncertainties $\delta\qasy$ and $\delta\Rasy$, respectively.
\\
From Fig.~\ref{fig:sasys} it is clear that the large errors do not allow to make any precise prediction on the asymptotic value of the ratio. It is interesting to notice how the convergence $|\GEq|\to|\GMq|$ as $q^2\to\infty$ is compatible with any case, apart from the one with $(\nth,\nasy)=(-1,2)$. Such a case is described by the darker green band, which however corresponds to the most probable one, see Fig.~\ref{fig:cases} and Table~\ref{tab:cases}.
\\
This also means that the knowledge of the modulus of the ratio at higher time-like $q^2$ would play a key role in disentangling the phase determination and hence in giving a hint on the pair $(\nth,\nasy)$ that describes the phase evolution with $q^2$, only in the case of $\Rasy>6$, implying $(\nth,\nasy)=(-1,2)$.
\\
On the other hand, a set of precise data on both the modulus of the ratio and its phase, preferably covering a quite wide range of $q^2$, would be certainly crucial in further reducing the possible cases, exactly as those already available have done in selecting from all possible pairs the only six that we have considered.
\subsection{The phase of $\GE/\GM$}
\label{subsec:phase}
As stated above, experiments measure the sinus of the phase $\arg\lt\GE/\GM\rt$, not giving a direct determination of the phase. This is not a pure mathematical issue, as it encodes precise information about the space-like behavior of the analytic function under consideration, the ratio $\GE/\GM$ in our case.
\\
It is interesting, however, to notice that, even though a direct knowledge of the phase determination is prevented by the periodicity of the sinus function, it turns out to be somehow accessible by looking at the $q^2$-trend, i.e., the evolution of the sinus function itself as $q^2$ increases from the theoretical threshold \qth\ on. Indeed, having determined the starting value at \qth, by means of the dispersive analysis, as one of the three possibilities: $\arg\lt\GE(\qth)/\GM (\qth)\rt=-\pi,0,\pi$, i.e., $\nth=-1,0,1$, the determination at a certain $q^2$ is given by the number of oscillations that the sinus function has undertaken from \qth\ up to that value of $q^2$. 
\begin{figure}[htb]	
\includegraphics[width=\columnwidth]{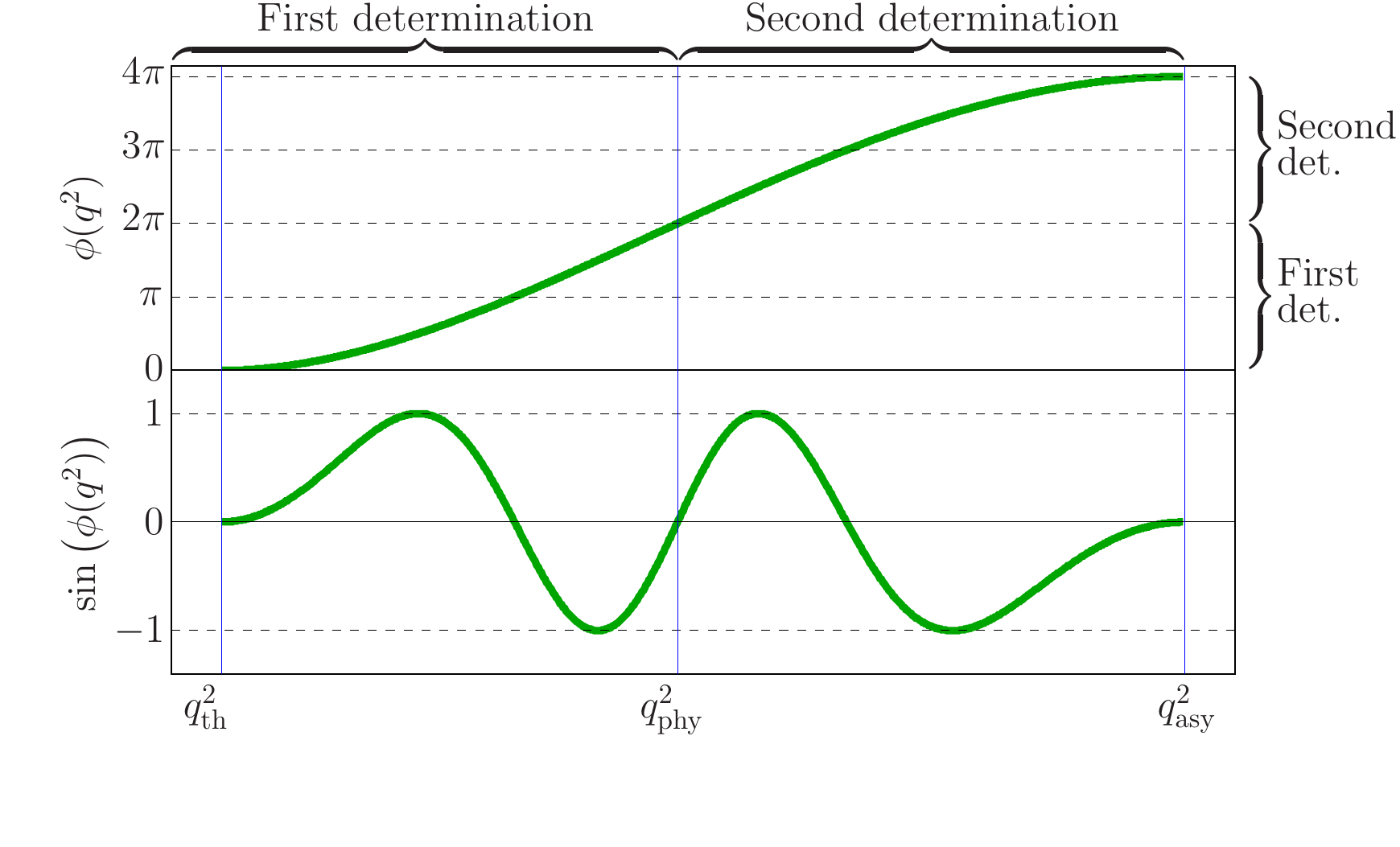}
\caption{\label{fig:example}The sinus of the phase $\phi(q^2)$, lower panel, undertakes two complete oscillations, having nodes at each of the three thresholds: \qth, \qphy\ and \qasy. In correspondence of such behavior of its sinus, the phase, upper panel, increases continuously, spanning the first and the second determination. It starts from zero at \qth, reaching $2\pi$ radians at \qphy\ and $4\pi$ radians at the asymptotic threshold \qasy.}
\end{figure}
\begin{figure}[htb]	
\begin{center}
\includegraphics[width=.49\columnwidth]{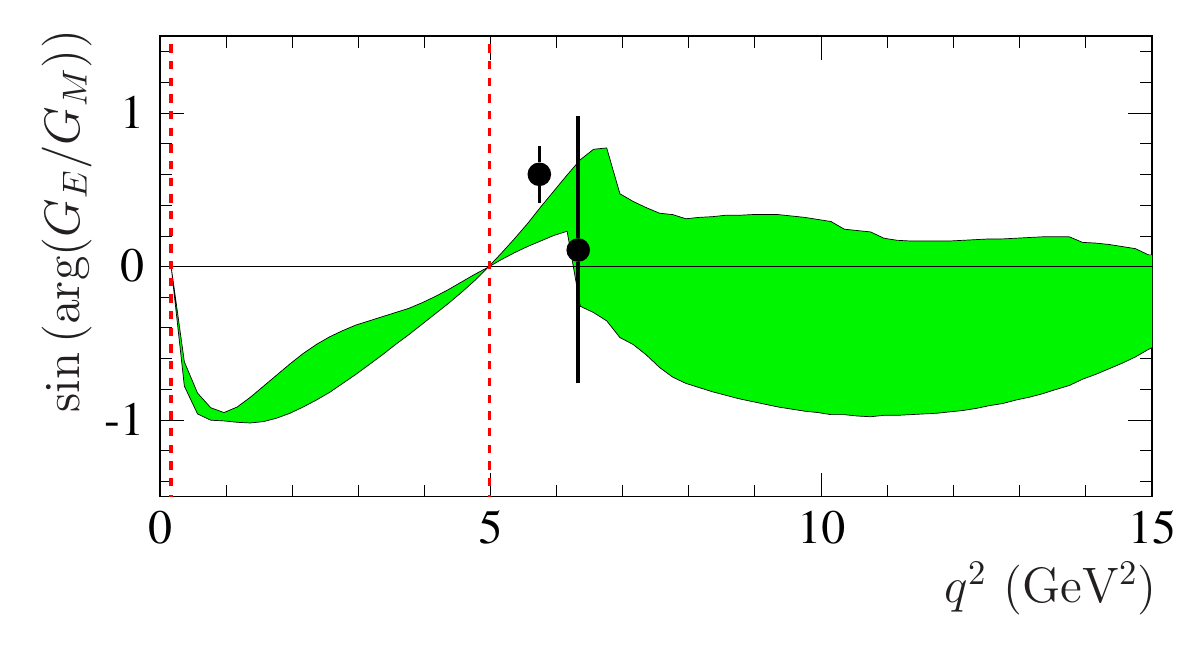}	
\includegraphics[width=.49\columnwidth]{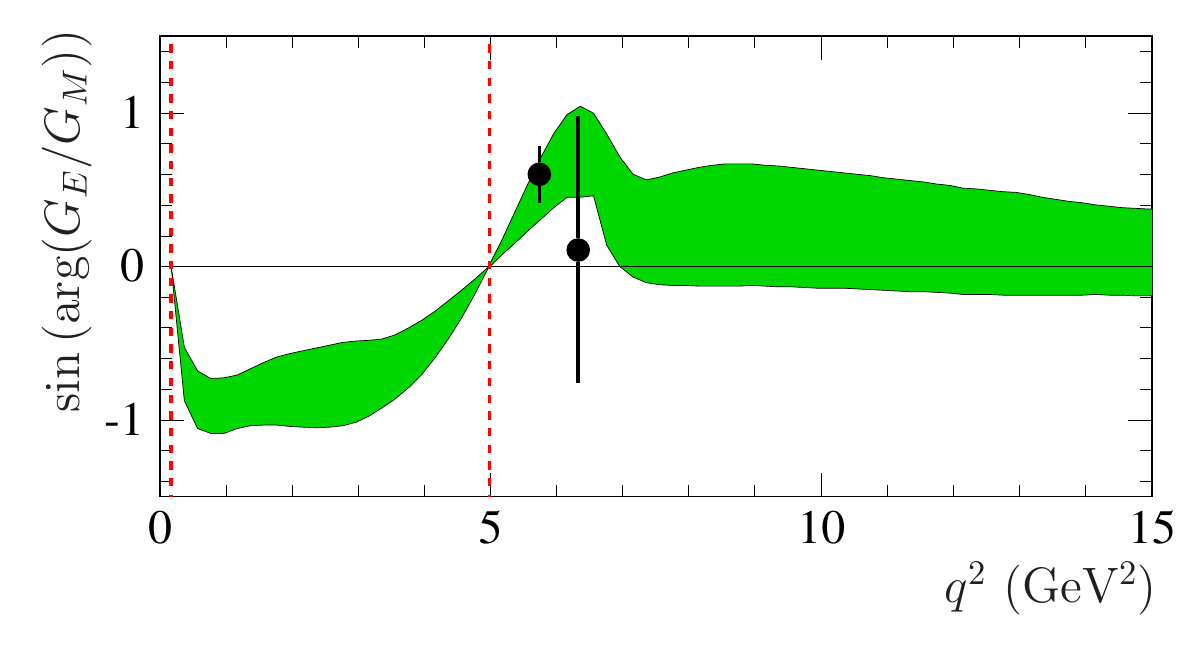}	
\includegraphics[width=.49\columnwidth]{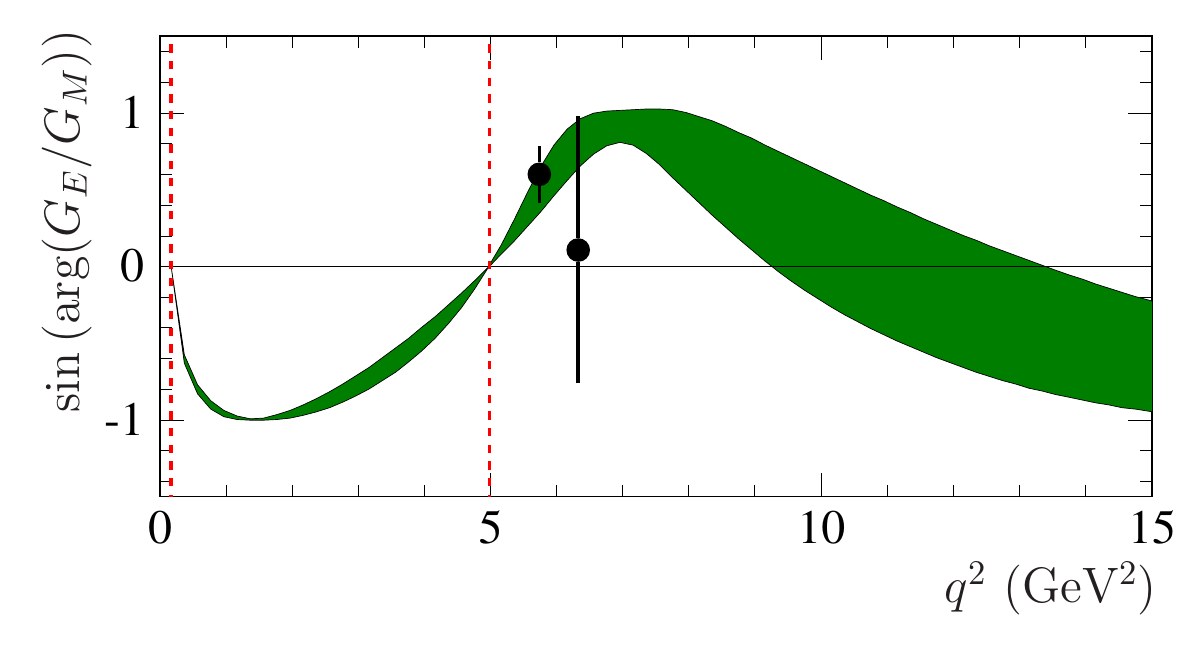}	
\includegraphics[width=.49\columnwidth]{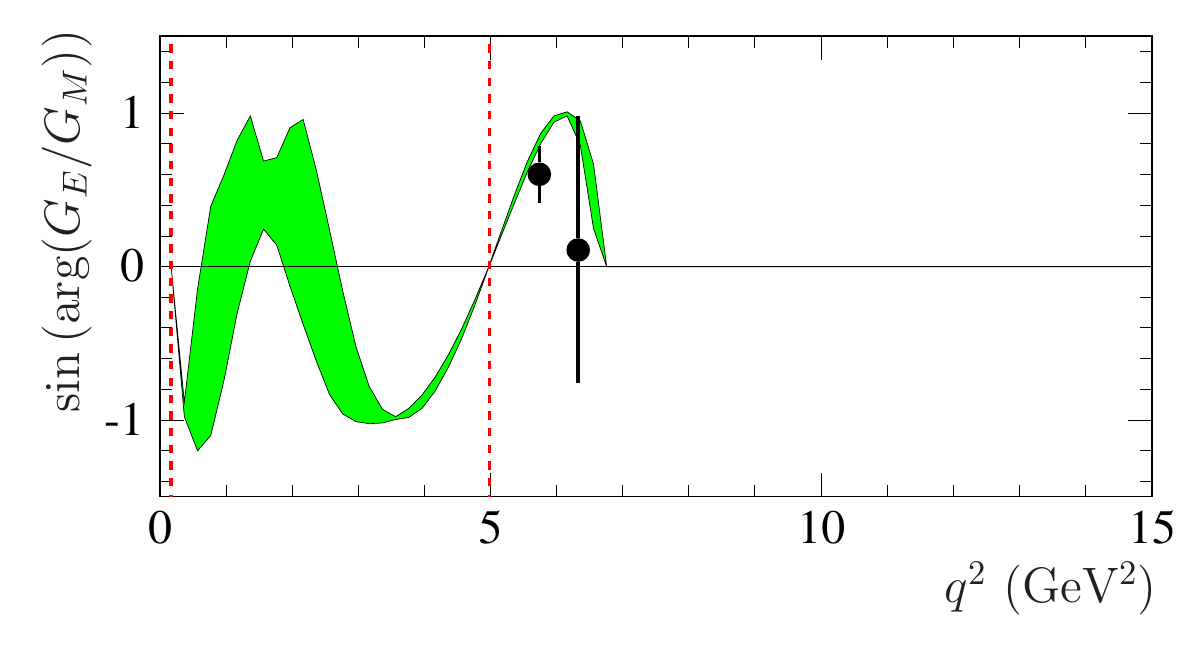}	
\includegraphics[width=.49\columnwidth]{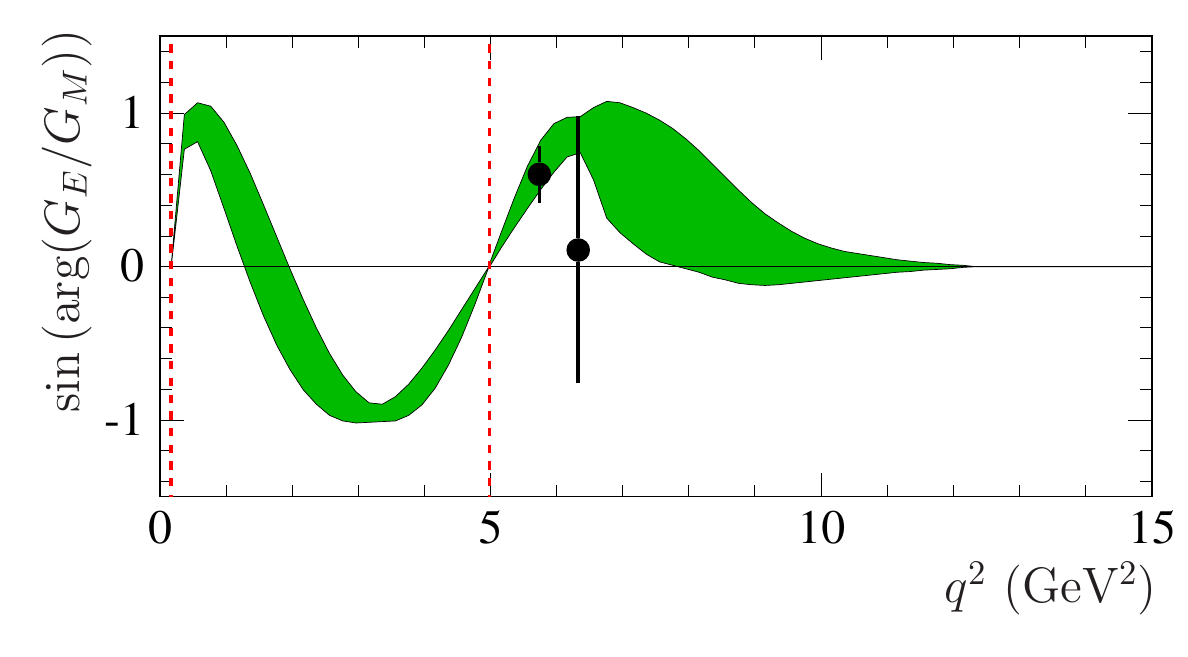}	
\includegraphics[width=.49\columnwidth]{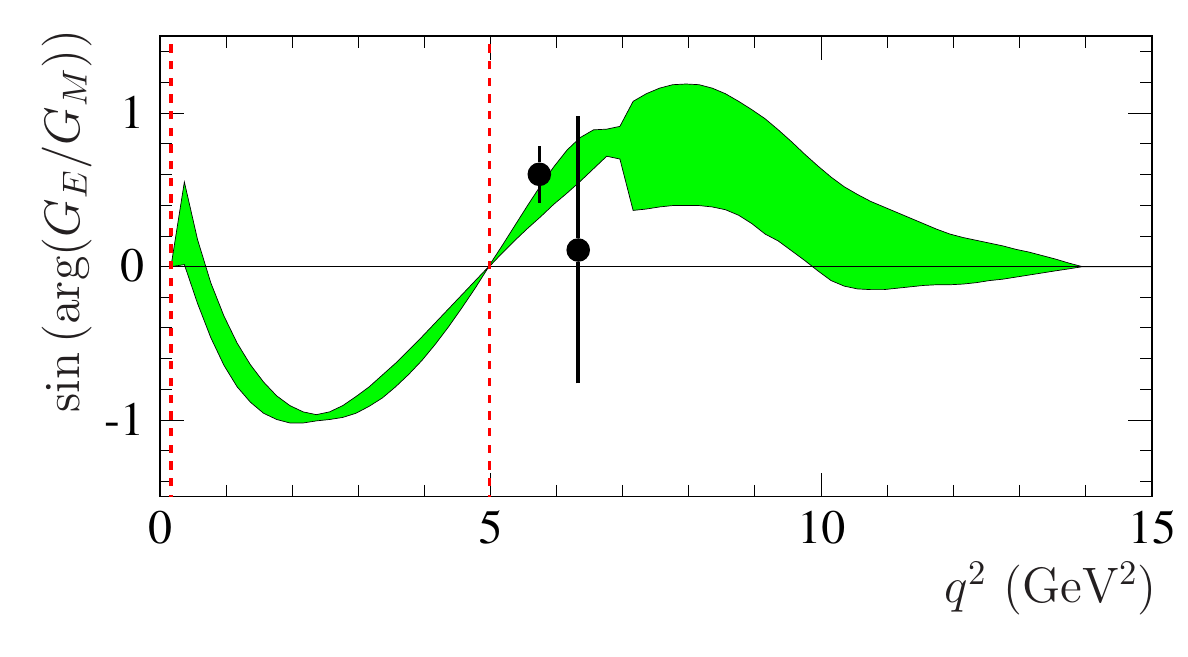}	
\caption{\label{fig:sinus}Sinus of the phase of the FF ratio with: $(\nth,\nasy)=(-1,0)$ in position left-up, $(\nth,\nasy)=(-1,1)$ in position right-up, $(\nth,\nasy)=(-1,2)$ in position left-middle, $(\nth,\nasy)=(-1,3)$ in position right-middle, $(\nth,\nasy)=(0,3)$ in position left-down, $(\nth,\nasy)=(1,3)$ in position right-down. The color intensity of the error bands is proportional to the occurrence probability of the corresponding pair $(\nth,\nasy)$, a darker color indicates an higher probability. The red dashed lines indicate the theoretical and the physical thresholds.
}\end{center}
\end{figure}\noindent
Figure~\ref{fig:example} shows just an example with $(\nth,\nasy)=(0,4)$, where, indeed, a given phase $\phi(q^2)$, upper panel, increases with $q^2$, starting from zero at the theoretical threshold, passing through $2\pi$ radians at the physical threshold, where the ratio becomes again real and finally reaching the asymptotic regime at \qasy, at which the ratio goes back to be real. From the experimental point of view, the data on the sinus of such a phase should follow the oscillating behavior shown in the lower panel of Fig.~\ref{fig:example}, i.e., it should describe a complete period, reaching the maximum value $+1$ and the minimum $-1$. This behavior, once the phase value at \qphy\ is determined by the dispersive analysis, would unambiguously reveal the determination of the phase and hence the resulting space-like features of the ratio. The message is that a measure covering a $q^2$ range as wide as possible would allow getting information about the phase with its own physical determination. 
The panels of Fig.~\ref{fig:sinus} represent the results for the sinus of the phase $\arg\lt\GE/\GM\rt$ obtained in the six cases under study, namely those with $\lt\nth,\nasy\rt=(-1,0),\,(-1,1),\,(-1,2),\,(-1,3),\,(0,3),\,(1,3)$. Following the scheme of the previous pictures, the level of darkness of the green areas, describing the error bands, is proportional to the occurrence probability of the corresponding case. Unfortunately, we notice that in the results characterized by large values of $\Delta N=\nasy-\nth$, most of the oscillations of the sinus of the phase lie below the physical threshold and hence are not experimentally observable. Nevertheless, a better knowledge of this quantity, that is, more precise data and covering a larger energy region would be a pivotal step forward in the understanding of dynamics underlying the $\gamma\Lambda\bar\Lambda$ interaction.
\subsection{The $\Lambda$ charge radius}
\label{subsec:radius}
The complete knowledge of the FF ratio $R(q^2)$, as an analytic complex function of $q^2$, defined in the whole $q^2$-complex plane, allows us to infer dynamical and also static features of the \L\ baryons.
\\
One example is the so-called charge root-mean-square radius, \rae, i.e., the square root of the mean square radius of the spatial charge distribution. In general, for an extended particle as a baryon, which have an electric FF $G_E(q^2)$, such a quantity is defined by
\be
\rae^2=\left.6\frac{dG_E(q^2)}{dq^2}\right|_{q^2=0}\,,
\nen
i.e., it is proportional to the first derivative of the electric FF with respect to the four-momentum squared at $q^2=0$. The definition follows from the fact that in the Breit Frame, being no energy exchange between electron and baryon, the four-momentum $q$ is purely space-like, i.e., $q=(0,\vec{q})$, then the electric FF represents the Fourier transform of the spacial charge distribution of the baryon.    
\\
It is interesting to notice that in the case of a neutral baryon with non-vanishing magnetic moment $\mu$, so that it has electric and magnetic FFs normalized as 
\be
G_E(0)=0\,,\hh G_M(0)=\mu\not=0\,,
\nen
the charge radius can be defined also in terms of the first derivative of the ratio  $R(q^2)=G_E(q^2)/G_M(q^2)$ always at $q^2=0$. In fact
\be
\left.\frac{dR(q^2)}{dq^2}\right|_{q^2=0}\ug 
\frac{1}{G_M(q^2)}\lt \frac{dG_E(q^2)}{dq^2}
\right.\no\\
&&\left.\left.-\frac{G_E(q^2)}{G_M(q^2)}\frac{d G_M(q^2)}{dq^2}\rt\right|_{q^2=0}\,,
\no\\\ug
\lt \frac{1}{G_M(q^2)}\left. \frac{dG_E(q^2)}{dq^2}\rt\right|_{q^2=0}
=\frac{1}{\mu}\frac{\rae^2}{6}\,,
\nen
it follows that
\be
\rae^2=\left.6\mu\frac{dR(q^2)}{dq^2}\right|_{q^2=0}\,.	
\nen
On the other hand, a DR for the first derivative of $R(q^2)$ can be obtained from that of Eq.~\eqref{eq:dr-im}, in particular, with $q^2<\qth$,
\be
\frac{dR(q^2)}{dq^2}=\frac{1}{\pi}\int_{\qth}^\infty\frac{\im\lt R(s)\rt}{(s-q^2)^2}ds\,,
\nen
that, evaluated at $q^2=0$ gives
\be
\left.\frac{dR(q^2)}{dq^2}\right|_{q^2=0}=\frac{1}{\pi}\int_{\qth}^\infty\frac{\im\lt R(s)\rt}{s^2}ds\,.
\nen
By considering the parameterization of Eq.~\eqref{eq:cheby}, this derivative can be expressed directly in terms of the free parameters of the set $\{ \vec{C}=\lt C_0,C_1,\ldots,C_N\rt,\qasy \}$ as it follows
\be
\left.\frac{dR(q^2)}{dq^2}\right|_{q^2=0}\!\!=
\frac{1}{\pi \Delta q^2}
\sum_{j=0}^N C_j\!\!\!\int_{-1}^{1}\!\!
\frac{T_j\lt x\rt}{\lt x+1+\qth/ \Delta q^2 \rt^2}dx\,\,\,\,,
\nen
with $\Delta q^2=(\qasy-\qth)/2$. 
\\
The magnetic moment of the \L\ baryon is $\mu_\L=-0.613\pm  0.004\,\mu_N$~\cite{pdg}, and the squared charge radius can be obtained as
\be
\rael^2=
\!\!=
\frac{6\mu_\L}{\pi \Delta q^2}
\sum_{j=0}^N C_j\!\!\!\int_{-1}^{1}\!\!
\frac{T_j\lt x\rt}{\lt x+1+\qth/ \Delta q^2 \rt^2}dx\,.\,\,\,
\nen
Depending on the set of free parameters, i.e., from the analysis results, this quantity can be both positive and negative. 
\\
The best known particle with a negative squared charge radius is the neutron. Such a feature can be explained by considering the electric charge spatial distribution of the neutron due to the valence quarks dislocation, which, assuming the spherical symmetry, can be further simplified in terms of their mean distance from the center of the neutron. The squared charge radius will be negative when the two {\it down} valence quarks, bearing the negative charge, lie at a mean distance from the neutron centre larger than the distance of the positively charged {\it up} quark. 
\\
 This unbalanced distribution of charge within the volume of the neutron gives rise to the small negative squared charge radius $\langle r_E^n\rangle^2=-0.1161\pm 0.0022$~fm$^2$~\cite{pdg}. In order to have a better understanding of the linear extension of the baryon, we define a normalized charge radius, by taking the square root of the modulus of the squared charge radius to which we assign the original sign, i.e.,
  \be
\bar r_E ={\rm Sign}\lq\rae^2\rq\,\sqrt{\left|\rae^2\right|}\,.
 \nen
 Following this definition, the normalized charge radius of the neutron is $\bar r_N^n=-0.3407\pm0.0032$ fm.
\begin{figure}[H]
		\includegraphics[width=\columnwidth]{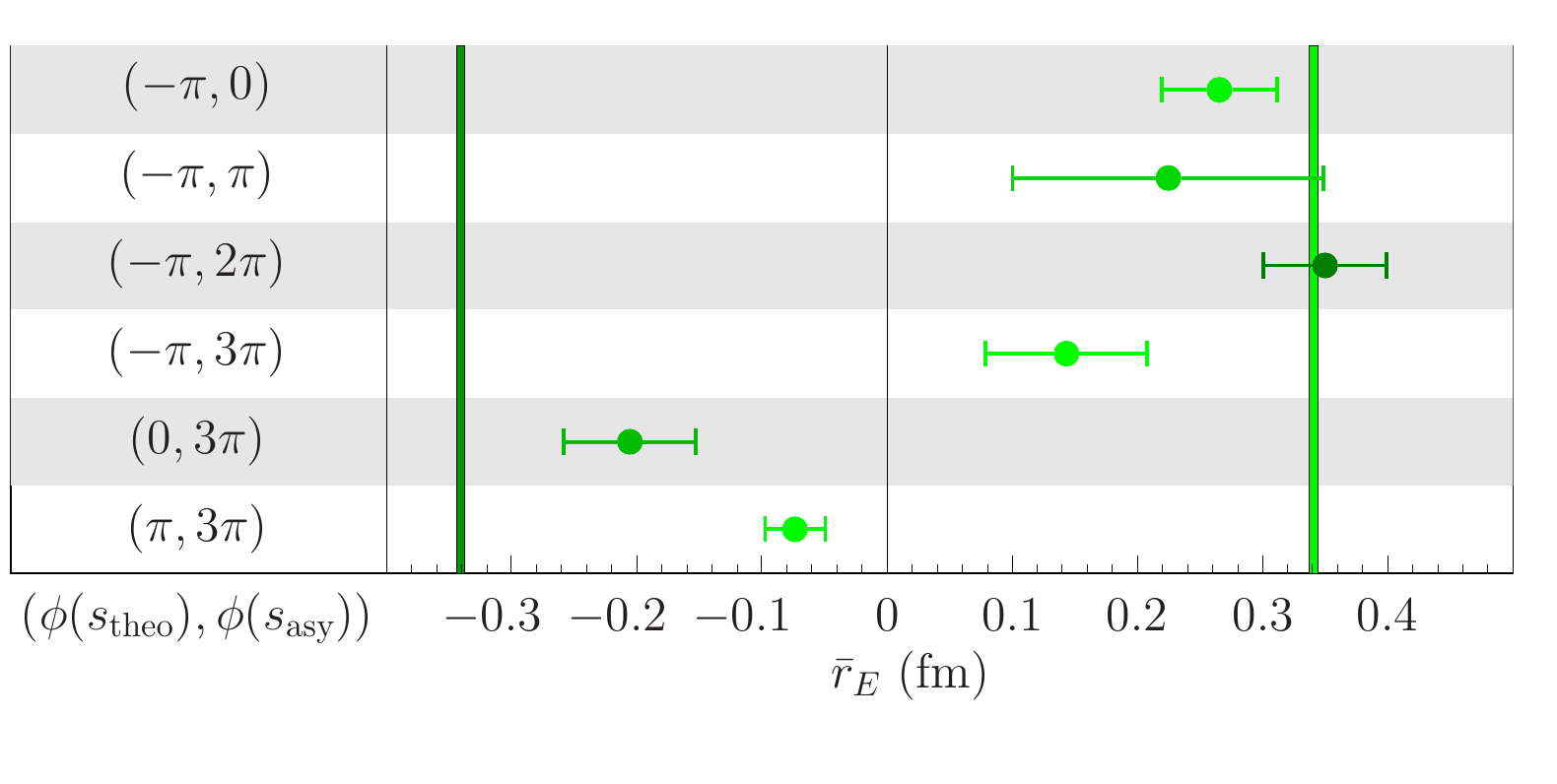}
		\caption{\label{fig:cases-radiii}The points represent the normalized charge radii obtained in the six cases under consideration, whose $(\nth,\nasy)$ pairs are reported on the left hand side of the picture. Also here, the intensity of the green color is proportional to the probability of occurrence of each $(\nth,\nasy)$ pair. For comparison, the symmetric vertical green bars indicate the negative normalized neutron charge radius, on the left, and its reflection on the right.}
\end{figure}\noindent
Figure~\ref{fig:cases-radiii} shows the quite different values of the normalized charge radius of the $\Lambda$ baryon obtained in this study, in comparison with that of the neutron, which is represented by the vertical dark green bar. For comparison, we also report its positive opposite, the light-green bar on the right hand side of Fig.~\ref{fig:cases-radiii}. 
\\
The heuristic interpretation of this result is that, while in the case of neutron the two {\it down} quarks orbit at a distance that in average is larger than that of the {\it up} quark, giving an overall effect of a negative squared charge radius, for the \L\ baryon, in the most probable case with $(\nth,\nasy)=(-1,2)$, see Table~\ref{tab:cases}, the configuration is the opposite. In fact, the negative charge, carried by the {\it down} and the heavy {\it strange} quarks, is mainly concentrated at small distances, resulting in a positive squared charge radius.   
\\
On the other hand, the second most probable result, corresponding to $(\nth,\nasy)=(0,3)$ and occurring with a probability of about half of that of the case with $(\nth,\nasy)=(-1,2)$, corresponds to a negative squared charge radius compatible with the that of the neutron. 
\\
Despite the powerful role played by the theoretical approach in establishing a set of reliable conditions which restrict the choice of possible configurations, the present discussion on the radius points to the need of further experimental investigation.
\section{Conclusion}
\label{sec:conlcusion}
In order to study the \L\ baryon electromagnetic FFs, we have defined a phenomenological approach, which is based on first principles as analyticity and optical theorem, that have been implemented by means of the powerful mathematical tool of DRs.  
\\
The first and unique precise experimental data of the phase of the FF ratio $\GE/\GM$, obtained by the BESIII Collaboration~\cite{Ablikim:2019vaj}, have been analyzed together with those on its moduli to gain a deeper knowledge of the dynamics which rules the electromagnetic coupling between the photon and the \LL\ system in the time-like region, as well as the space-like counter-part, i.e., the amplitude of the radiative virtual conversion $\L(\bar \L)\to\gamma\bar\L(\L)$. 
\\
This is, actually, the key point of our study, that is, the connection between the time-like and the space-like behaviors of the \L\ FFs. In particular, we take advantage of the fact, formalized by the Levinson's theorem~\cite{levinson}, that the variation of the phase as a function of $q^2$ in the time-like interval, going from the theoretical threshold \qth\ up to infinity, is directly connected to the number of zeros of the electric FF \GE\ in the space-like region. This kinematic region, characterized by negative values of $q^2$, is not experimentally accessible for the \L\ baryon as well as for all other heavier baryons, because their instability prevents scattering experiments with electron or muon beams.
\\
The connection between time-like and space-like behaviors provided by this dispersive approach represents a unique possibility to investigate the FFs at negative $q^2$ and then gain information on dynamical and static properties of the baryons. In more detail, concerning the phase, we find that in each of the six cases, reported in Table~\ref{tab:cases}, which, in the framework of our phenomenological description, are compatible with the exiguous set of available data, there is always a non vanishing phase variation, i.e., 
\be
\Delta\phi=\phi(\infty)-\phi(\qth)
=\Delta N\pi=(\nasy-\nth)\pi
\ge\pi\,.
\nen
This means that the electric FF has at least one zero in the $q^2$ region below the theoretical threshold \qth and confirms the time-like $\leftrightarrow$ space-like connection. In fact, the electric FF, being the \L\ a neutral baryon, must be normalized to zero at $q^2=0$.
\\
The fact that the most probable solutions, see Table~\ref{tab:cases}, are those with $\Delta\phi=3\pi$, with $(\nth,\nasy)=(-1,2)$ and $(0,3)$, represents a clear indication that \GE\ has additional zeros. Such an eventuality, if confirmed by further studies and more exhaustive experimental investigations, would be a quite interesting feature that could be interpreted in terms of charge distribution and hence of the dynamical mechanism underlying the electromagnetic coupling of the \L\ baryons.
\\
The presence of unexpected space-like zeros for the proton electric FF is object of a deep theoretical and experimental investigation since the years 2000 at MIT Bates and at Jefferson Lab (see Ref.~\cite{Puckett:2017flj} and references therin). Here the first measurements of the proton FF ratio $G_E^p/G_M^p$ have been performed by exploiting, for the first time, the polarization-transfer technique~\cite{Akhiezer:1968ek,Akhiezer:1973xbf}, obtaining results compatible with the presence of a zero for the ratio, and hence for $G_E^p$, lying at $q^2\sim -10$ GeV$^2$. The results of the JLab-GEp Collaboration give only a hint for a space-like zero, as they show a monotone decrease of the proton FF ratio, that, when extrapolated, would cross zero. The planned experiments at Jefferson Lab~\cite{new-JLab3}, by extending the kinematical region up to $q^2=-15$ GeV$^2$, will answer this question. Once again, a space-like zero for the proton electric FF determines the time-like behavior of the phase of the ratio $G_E^p/G_M^p$, which should have an increase of $\pi$ radians from the theoretical up to the asymptotic threshold. However, since the proton is a stable baryon, the only possibility to obtain its polarization vector and hence to gain information on the $G_E^p/G_M^p$ phase, relies on a direct measurement of the polarization itself. Unfortunately, there are neither running nor planned future experiments able to perform such a kind of measurements in the time-like region.
\\
Finally, we have discussed the possibility of observing the determination of the phase, despite the fact that the measured observable, i.e., the sinus of the phase, does not contain this information. This severe limitation can be overcome by studying the $q^2$-evolution of the sinus of the phase in connection with the modulus of the FF ratio, in the framework of our dispersive procedure. In fact, as pointed out in Sec.~\ref{subsec:phase}, assuming that the ratio $R(q^2)=\GEq/\GMq$ is an analytic function in the whole $q^2$ complex plane, with the branch cut $(\qth,\infty)$, which means that the magnetic FF has no zeros in this domain, the phase can only increase as $q^2$ runs from the theoretical, \qth, up to the asymptotic threshold \qasy. 
\\
As an example, we can consider a phase increasing from the multiple \nth\ of $\pi$ radians, at \qth, to the multiple \nasy, at \qasy, as shown in the upper panel of Fig.~\ref{fig:example}, so that $(\nth,\nasy)=(0 , 4)$. To such a phase corresponds an oscillating behavior of the sinus, which, starting and ending up at zero, undertakes a number of oscillations equal to $\Delta N=\nasy-\nth$. Here an oscillation is defined as the portion of the sinus function passing through either a maximum $+1$ or a minimum $-1$, included between two consecutive nodes. The lower panel of Fig.~\ref{fig:example} shows the four oscillations, passing through two maxima and two minima, undertaken by the sinus of the phase, shown in the upper panel, which, indeed, increases from zero up to $4\pi$ radians.
\\
The actual experimental situation, even though is characterized by only one precise datum on the phase provided by the BESIII Collaboration, $\sin(\phi_{\rm BESIII})=0.60\pm 0.19$ at $\sqrt{q^2}=2.396$~GeV, appears quite promising. Indeed, this value of the sinus significantly different from zero is a clear indication that the phase at the corresponding $q^2$ is still increasing, and hence that the asymptotic threshold has been not yet reached. 
\\
In other words, the unique BESIII datum is sufficient to indicate that there is a non-negligible probability that, by exploring a wider $q^2$ region, more oscillations could be seen. The knowledge of the asymptotic value of the phase, i.e., the number of oscillations of its sinus would allow us to know how many zeros the electric FF has in the experimentally inaccessible space-like region. This is a crucial information on the internal structure of the \L\ baryon. 
\\
Finally, a second precise determination of the sinus of the phase at a different $q^2$, not too far from that of the BESIII datum, would allow to establish if the sinus at $\sqrt{q^2}=2.396$~GeV is either increasing or decreasing. While in the latter case we could expect that the phase is approaching its asymptotic threshold and hence what we are measuring is just the last oscillation of its sinus, in the former case, i.e., if the sinus is increasing, we have to expect a further complete oscillation.    
\section*{Acknowledgements}
It is a privilege to acknowledge Rinaldo Baldini Ferroli for being an invaluable and boundless source of brilliant new ideas which pour out from any conversation with him. 
\\
This work was supported in part by the STRONG-2020 project of the European Union's Horizon 2020 research and innovation program under Grant agreement number 824093.
%
%
%
%
%
%
%
%
%
%
%
%
%
%
%
%
\appendix
\section{Schemes of QCD corrections}
\label{app:qcd-corr}
For illustration, we consider the following two examples where the implementation of QCD corrections leads to different asymptotic behaviors for the FF ratio $R(q^2)$.
\begin{enumerate}
\item From their analysis, performed in the framework of perturbative QCD, A. V. Belitsky, X. Ji, F. Yuan~\cite{Belitsky:2002kj} obtained for the ratio of the Dirac, $F_1$, and Pauli FF, $F_2$, at the logarithmic accuracy, the asymptotic behavior
\be
\frac{F_2(q^2)}{F_1(q^2)}\mathop{\propto}_{|q^2|\to\infty}-\frac{\ln^2(q^2/\Lambda_{\rm soft}^2)}{q^2}
\label{eq:asy-f1-f2}
\nen
where $\Lambda_{\rm soft}$ is what they called \emph{a soft scale related to the size of the baryon}\footnote{In natural units $1\,{\rm fm}^{-1}\simeq 197$ MeV.}, that is of the order of a few hundreds of MeV. In this case the ratio diverges like $\ln^2(|q^2|)$ as $|q^2|\to\infty$. Indeed, by exploiting the well-known expressions of the electric and magnetic FFs in terms of the Dirac and Pauli ones, i.e.,  
\be
R(q^2)\ug \frac{\GE(q^2)}{\GM(q^2)} 
\no\\\ug
\frac{F_1^\Lambda(q^2)+\frac{q^2}{4M_\Lambda^2}F_2^\Lambda(q^2)}{F_1^\Lambda(q^2)+F_2^\Lambda(q^2)}\no
\\
\ug
\frac{1+\frac{q^2}{4M_\Lambda^2}\frac{F_2^\Lambda(q^2)}{F_1^\Lambda(q^2)}}{1+\frac{F_2^\Lambda(q^2)}{F_1^\Lambda(q^2)}}\,,
\label{eq:r-f1f2}
\en
and using the result of Eq.~\eqref{eq:asy-f1-f2} for the ratio $F^\Lambda_2/F^\Lambda_1$, we have
\be
R(q^2)\mathop{\propto}_{|q^2|\to\infty}-
\frac{\ln^2(q^2/\Lambda_{\rm soft}^2)}{4M_\Lambda^2}\,.
\label{eq:log-div}
\en
\item The second example is due to M.~Gari and W.~Krumpelmann~\cite{Gari:1984rr,Gari:1984ia,Gari:1986rj} who proposed to consider QCD correction by substituting $q^2$ with 
\be
\ \ \ \ \ \ \
q^2\to \tilde q^2=q^2\frac{\ln\lt \Lambda_2^2-q^2\rt -\ln\lt \Lambda_{\rm QCD}^2\rt}{\ln\lt\Lambda_2^2\rt -\ln\lt \Lambda_{\rm QCD}^2\rt}\,,
\nen
where $\Lambda_{\rm QCD}= 0.3$~GeV, while $\Lambda_2$, being of the order of a few GeV, represents the threshold from which FFs start to follow the perturbative QCD power-law. In this case the ratio of the Dirac and Pauli FFs scales like
\be
 \frac{F_2(q^2)}{F_1(q^2)}
\mathop{\sim}_{|q^2|\to\infty}
-\frac{[{\rm constant}]}{q^2\ln\lt q^2/\Lambda_{\rm QCD}^2\rt}\,.
\label{eq:asy-f1-f2-2}
\nen
It follows that the ratio of FFs is asymptotically constant and equal to one, i.e.,
\be
\ \ \ \ \ \ \ \ \
R(q^2)\mathop{\propto}_{|q^2|\to\infty}
\frac{1-\frac{[{\rm constant}]}{4M_\Lambda^2\ln\lt q^2/\Lambda_{\rm QCD}^2\rt}
}{1-\frac{[{\rm constant}]}{q^2\ln\lt q^2/\Lambda_{\rm QCD}^2\rt}}\mathop{\to}_{|q^2|\to \infty}1\,,
\nen
where we have used the expression of the ratio given in Eq.~\eqref{eq:r-f1f2}.
\end{enumerate}
%
%
%
%
%
%
%
%
%
%
%
%
%

\end{document}